\documentclass[preprint,aps,eqsecnum,floatfix,showpacs]{revtex4-1}

\def\Authors{L.\ Ceccarelli\,$^{1}$, A.\ Gnech\,$^{2}$, L.E.\ Marcucci\,$^{1,3,*}$, M.\ Piarulli\,$^{4,5}$ and M.\ Viviani\,$^{3}$}
\def\Address{$^{1}$Dipartimento di Fisica ``E.\ Fermi'', Universit\`a di Pisa,
  Pisa I-56127, Italy \\
  $^{2}$ Theory Center, Jefferson Lab, Newport News, Virginia 23606, USA \\
  $^{3}$Istituto Nazionale di Fisica Nucleare, Sezione di Pisa,
  Pisa I-56127, Italy \\
  $^{4}$ Physics Department, Washington University, St.\ Louis, MO 63130, USA \\
  $^{5}$ McDonnell Center for the Space Sciences at Washington University in
  St.\ Louis, MO 63130, USA}

\usepackage{bm}
\usepackage{color}
\usepackage{graphicx}
\usepackage{pstricks}

\begin{document}

\title{Muon capture on deuteron using local chiral potentials} 

\author{\Authors} 
\affiliation{\Address} 

\begin{abstract}
The muon capture reaction
$\mu^- + d \rightarrow n + n + \nu_{\mu}$
in the doublet hyperfine
state is studied using nuclear potentials and consistent
currents derived in chiral effective field theory, which are local and
expressed in coordinate-space (the so-called Norfolk models).
Only the largest contribution due to the
$^1S_0$ $nn$ scattering state is considered. Particular attention is given
to the estimate of the theoretical uncertainty, for which
four sources have been identified:
(i) the model dependence,
(ii) the chiral order convergence for the weak nuclear current,
(iii) the uncertainty in the single-nucleon axial form factor,
and
(iv) the numerical technique adopted to solve the bound and scattering
$A=2$ systems. This last source of uncertainty has turned out
essentially negligible. 
The $^1S_0$ doublet muon capture rate $\Gamma^D(^1S_0)$
has been found to be $\Gamma^D(^1S_0)=255.8(0.6)(4.4)(2.9)$ s$^{-1}$,
where the three errors
come from the first three sources of uncertainty.
The value for $\Gamma^D(^1S_0)$ obtained within this local
chiral framework is compared with previous calculations
and found in very good agreement.
\end{abstract}

\keywords{Muon capture, deuteron, chiral effective
  field theory, ab-initio calculation, error estimate} 
\maketitle

\section{Introduction}
\label{sec:intro}

The muon capture on deuteron, i.e.\ the process 
\begin{equation}
\mu^- + d \rightarrow n + n + \nu_{\mu}\label{reaction} \ ,
\end{equation}
is one of the few weak nuclear reactions involving light nuclei
which, on one side, are experimentally accessible,
and, on the other, can be studied using 
{\it ab-initio} methods. Furthermore, it is a process closely linked
to the proton-proton weak capture, the so-called $pp$ reaction,
\begin{equation}
p + p \rightarrow d + e^+ + \nu_{e} \label{reaction2} \ ,
\end{equation}
which, although being of paramount importance in astrophysics, is
not experimentally accessible, due to its
extremely low rate, and can only be calculated. Since the
theoretical inputs  to study reaction~(\ref{reaction2}) and
reaction~(\ref{reaction}) are essentially the same, the comparison
between experiment and theory for muon capture provides
a strong test for the $pp$ studies. 

The muon capture reaction~(\ref{reaction})
can take place in two different hyperfine states, $f=1/2$ and $3/2$.
Since it is well known that the doublet capture rate is about
40 times larger than the quartet one
(see for instance Ref.~\cite{Measday2001}),
we will consider the $f=1/2$ state only, and we will focus on the
doublet capture rate, $\Gamma^D$.

The experimental situation for $\Gamma^D$ is
quite confused, with available measurements which are relatively old. These are
the ones of Refs.~\cite{Wang1965,Bertin1973,Bardin1986,Cargnelli1989},
$365(96)\;\rm{s}^{-1}$, $445(60)\;\rm{s}^{-1}$, $470(29)\;\rm{s}^{-1}$ and
$409(40)\;\rm{s}^{-1}$, respectively.
All these data are consistent with each other within the experimental
uncertainties, which are however quite large. In order to clarify the
situation, an experiment with the aim of measuring $\Gamma^D$ with a 1\%
accuracy is currently performed at the Paul Scherrer Institute,
in Switzerland, by the MuSun Collaboration \cite{Kammel2012}.

Many theoretical studies are available for the muon capture rate
$\Gamma^D$. A review of the available literature of 
up to about ten years ago can be found in Ref.~\cite{Marcucci2012review}.
Here we focus on the work done in the past ten years. To the best
of our knowledge, the capture rate $\Gamma^D$ has been studied in
Refs.~\cite{Adam2011,Marcucci2011,Marcucci2012chiral,Golak2014,Acharya2018}.
The studies of Refs.~\cite{Marcucci2011,Golak2014} have been performed
within the phenomenological approach, using phenomenological potentials
and currents. In Ref.~\cite{Marcucci2011}, the first attempt
to use chiral effective field theory ($\chi$EFT) was presented,
within the so-called  
hybrid approach, where a phenomenological nuclear interaction is used
in conjunction with $\chi$EFT weak nuclear charge and current operators. In
the study we present in this contribution, though, we are interested
not only in the
determination of $\Gamma^D$, but also in an assessment of the theoretical
uncertainty. This can be grasped more comfortably and robustly
within a consistent
$\chi$EFT approach. Therefore, we review only the
theoretical works of Refs.~\cite{Adam2011,Marcucci2012chiral,Acharya2018}, which have been
performed within a consistent $\chi$EFT.
The studies of Refs.~\cite{Adam2011} and~\cite{Marcucci2012chiral} were essentially
performed in parallel. They both employed the latest (at those times)
nuclear chiral potentials and consistent weak current operators.
In Ref.~\cite{Adam2011}, the doublet capture rate was found to be 
$\Gamma^D=388.1(4.3)$ s$^{-1}$, when the $NN$ chiral potentials
of Ref.~\cite{Entem2003}, obtained up to next-to-next-to-next-to leading
order (N3LO) in the chiral expansion, were used.  When only the $^1S_0$ channel
of the final $nn$ scattering
state was retained, it was found $\Gamma^D(^1S_0)=247.7(2.8)$ s$^{-1}$.
In Ref.~\cite{Marcucci2012chiral}, a simultaneous study of the muon
capture on deuteron and $^3$He was perfomed using the same
N3LO chiral potentials, but varying the potential cutoff
$\Lambda=500,600$ MeV~\cite{Entem2003,Machleidt2011}, and consequently
refitting consistently for each value of $\Lambda$
the low-energy constants (LECs)
entering into the axial and vector current operators.
For the muon capture on deuteron, it was obtained
$\Gamma^D=399(3)$ s$^{-1}$, the spread accounting
for the cutoff sensitivity, as well as uncertainties in the LECs
and electroweak radiative corrections. When only the
$^1S_0$ channel is considered, $\Gamma^D(^1S_0)=254.9(1.4)$ s$^{-1}$,
where, in this case, the (small) uncertainty arising from
electroweak radiative corrections is not included. 
In the case of
the muon capture on $^3$He,
an excellent agreement with the available extremely accurate
experimental datum was found. Although obtained by different
groups and with some differences
in the axial and vector current operators adopted in the
calculations, the results of Refs.~\cite{Adam2011} and~\cite{Marcucci2012chiral}
for $\Gamma^D$ and $\Gamma^D(^1S_0)$ should be
considered in reasonable agreement.
It should be mentioned that in both studies of Refs.~\cite{Adam2011}
and~\cite{Marcucci2012chiral}, a relation between the LEC entering the axial
current operator (denoted with $d_R$)
and $c_D$, one of the two LECs entering the three-nucleon
potential (the other one being $c_E$)
was taken from Ref.~\cite{Gazit2009}. Then, the
$A=3$ binding energies and the Gamow-Teller of the triton $\beta$-decay
were used to fix both $c_D$ (and consequently
$d_R$) and $c_E$ for each given potential and cutoff $\Lambda$. Unfortunately,
the relation between $d_R$ and $c_D$ of
Ref.~\cite{Gazit2009} has been found to be missing of a
factor $-1/4$, as clearly stated in the Erratum of
Ref.~\cite{Marcucci2012chiral}
(see also the Erratum of Ref.~\cite{Gazit2009}). While the work
of Ref.~\cite{Adam2011} has not yet been revisited, that of
Ref.~\cite{Marcucci2012chiral} has been corrected, finding very
small changes in the final results, which become
$\Gamma^D=398(3)$ s$^{-1}$ and $\Gamma^D(^1S_0)=253.5(1.2)$ s$^{-1}$.

The most recent and systematic study of reaction~(\ref{reaction})
in $\chi$EFT,
even if only retaining the $^1S_0$ $nn$ channel, is that of
Ref.~\cite{Acharya2018}. There, $\Gamma^D(^1S_0)$ has been calculated
using a pool of 42
non-local chiral potentials up to next-to-next-to-leading order (N2LO),
with a regulator cutoff $\Lambda$ in the range 450-600 MeV
and six different energy ranges in the $NN$ scattering
database~\cite{Carlsson2016}.
The consistent axial and vector currents were constructed
(with the correct relation between $d_R$ and $c_D$),
and a simultaneous fitting procedure for all the involved LECs
was adopted. The final result was found to be
$\Gamma^D(^1S_0)=252.8(4.6)(3.9)\;\rm{s}^{-1}$,
in excellent agreement with Ref.~\cite{Marcucci2012chiral}.
Here the first error is due to the truncation in the
chiral expansion and the second one to 
the uncertainty in the parameterization of
the single-nucleon axial form factor (see below).
In Ref.~\cite{Acharya2018} it was also
questioned the accuracy of the variational method used to calculate
the deuteron and $nn$ scattering wave
functions in Refs.~\cite{Marcucci2011,Marcucci2012chiral}.
This same issue was already raised in Ref.~\cite{Acharya2017},
where it was found that a non-proper treatment of the infrared cutoff
when the bound-state wave function is represented in a truncated basis
(as in the case of Refs.~\cite{Marcucci2011,Marcucci2012chiral}) can lead
to an error of the order of $\sim 1$\% in the few-nucleon
capture cross sections and astrophysical $S$-factors
(as for instance that of the $pp$ reaction).

The chiral nuclear potentials involved in all the above mentioned
studies are highly non-local, and are expressed in momentum-space. This
is clearly less desirable compared with $r$-space in the case of
the $pp$ reaction, where the treatment in momentum-space of the Coulomb
interaction and of the higher-order electromagnetic effects is rather
cumbersome. In order to overcome these difficulties, local
chiral potentials expressed in $r$-space would be highly desirable. These
have been developed only in recent years,
as discussed in the recent review of Ref.~\cite{Piarulli2020}.
These potentials are very accurate, and have proven to be extremely
successful
in order to describe the structure and dynamics of light and
medium-mass nuclei. In particular,
we are interested in this work to the models of
Ref.~\cite{Piarulli2016}, the so-called Norfolk potentials,
for which, in these years, consistent electromagnetic and weak
transition operators have been
constructed~\cite{Baroni2018,Schiavilla2019,Gnech2022}.
This local chiral framework has been used to calculate energies~\cite{Piarulli2017}, charge radii~\cite{Gandolfi2020} and various electromagnetic observables in light nuclei, as the charge form factors in $A=6,12$~\cite{Gandolfi2020} and the magnetic structure of few-nucleon systems~\cite{Gnech2022}. It has been used also to study
weak transitions in light nuclei~\cite{King2020a,King2020b},
the muon captures on $A=3,6$ nuclei~\cite{King2022a}, neutrinoless double $\beta$-decay for $A=6,12$ ~\cite{Cirigliano2019} and the $\beta$-decay spectra in $A=6$~\cite{King2022b}, and, finally, also the equation of state of pure neutron matter~\cite{Piarulli2019,Lovato2022}. However, the use of the Norfolk potentials
to study the muon capture on deuteron~(\ref{reaction})
and the $pp$ reaction~(\ref{reaction2}) is still lacking.
It is one of
the aim of the present work to start this path. Given the fact that
$\Gamma^D(^1S_0)$ is the main contribution to $\Gamma^D$, and the $^1S_0$
channel is also the only one of interest for the $pp$
fusion~\cite{Acharya2019,Marcucci2013}, we focus here our attention
only on $\Gamma^D(^1S_0)$. A full calculation of $\Gamma^D$,
together with the rates for muon capture on $A=3$ and 6 nuclei,
is currently underway. The second aim of the present study is to
provide a more robust determination of the theoretical uncertainty
compared with the work of Ref.~\cite{Marcucci2012chiral}, although
probably not as robust as the full work presented in Ref.~\cite{Acharya2018}.
However, the procedure we plan to apply in the present work
is much simpler and, as it will be shown below, with a quite similar
final outcome. In fact, we will consider four sources
of uncertainties: (i) the first one is due to
model dependence. In this study, the use of the local Norfolk potentials
will allow us to take into consideration the uncertainty
arising from the cutoff variation, as well as the
energy ranges in the $NN$ scattering
database up to which the LECs are fitted. In fact,
as it will be explained in Sec.~\ref{subsec:pot-curr}, we will employ
four different versions of the Norfolk potentials, obtained
using two different sets of short- and long-range cutoffs, and
two different energy ranges, up to 125 MeV or up to 200 MeV, in the
$NN$ scattering database.
(ii) A second 
source of uncertainty arises from
the chiral order convergence. In principle, this should be
investigated by maintaining the same order for potentials
and weak nuclear currents. However, at present, the Norfolk potentials,
for which weak current operators have been consistently constructed,
are those obtained at N3LO. As a matter of fact, this chiral 
order is needed to reach a good 
accuracy in the description of the $NN$ systems and of light nuclei.
Therefore, it is questionable whether a study of
reaction~(\ref{reaction}) using potentials and currents at a chiral
order which do not even reproduce the nuclear systems under consideration,
would be of real interest. 
As a consequence, we will study in the present work only the chiral
order convergence for the weak nuclear currents, keeping fixed
the chiral order of the adopted potentials.
(iii) A third source of uncertainty is due to
the uncertainty in the
parameterization of single-nucleon axial form factor
$g_A(q_\sigma^2)$ as function of the squared four-momentum transfer $q_\sigma^2$.
This aspect will be discussed in details in Sec.~\ref{subsec:pot-curr}. Here
we only notice that the most recent parameterization for the
single-nucleon axial form factor is given by
\begin{equation}
g_A(q_{\sigma}^2)=g_A\left(1-\frac{1}{6}r_A^2q_{\sigma}^2+...\right)\label{esp}\ ,
\end{equation}
where the dots indicate higher-order terms, which are typically
disregarded, and $r_A$ is the axial charge radius, its square
being given by
$r_A^2=0.46(16)$ fm$^2$~\cite{Hill2018}. The large uncertainty
on $r_A^2$ will affect significantly the total
uncertainty budget, as already found in Ref.~\cite{Acharya2018}.
(iv) A final source of uncertainty is the one arising from
the numerical technique adopted to solve the bound and scattering
$A=2$ systems. In fact, taking into consideration the arguments
of Ref.~\cite{Acharya2017}, we have decided to use two methods. The
first one is the method
already developed in Refs.~\cite{Marcucci2011,Marcucci2012chiral},
i.e.\ a variational method, in which the bound and scattering
wave functions are expanded on a known basis, and the unknown
coefficients of these expansions are obtained by means of 
variational principles. The second method is the so-called Numerov method,
where the tail of the bound state wave function is in fact
imposed ``by hand'' (see Sec.~(\ref{subsec:nuclwf})).
This last source of uncertainty will be shown to be
completely negligible. This seems to be in contrast,
at least for the observable here under study, with the conclusions
of Ref.~\cite{Acharya2017}. 

The paper is organized as follows: in Sec.~\ref{sec:formalism}
we will present the theoretical formalism, providing a schematic
derivation for $\Gamma^D(^1S_0)$ in Sec.~\ref{subsec:obs},
a description 
of the adopted nuclear potentials and currents
in Sec.~\ref{subsec:pot-curr},
and a discussion of the
methods used to calculate the deuteron and $nn$ wave functions
in Sec.~\ref{subsec:nuclwf}.
The results for $\Gamma^D(^1S_0)$ will be presented and discussed
in Sec.~\ref{sec:results}, and some concluding remarks and an
outlook will be given in Sec.~\ref{sec:concl}.

\section{Theoretical formalism}
\label{sec:formalism}

We discuss in this section the theoretical formalism
developed to calculate the muon capture rate. In particular,
in Sec.~\ref{subsec:obs} we report the main steps
of the formalism used to derive the differential and the total
muon capture rate on deuteron in the initial 
doublet hyperfine state. A thourough discussion 
has been given in
Ref.~\cite{Marcucci2011}.
In Sec.~\ref{subsec:pot-curr} we report the main
characteristics of the nuclear potentials and currents
we have used in the present study. Finally in Sec.~\ref{subsec:nuclwf}
we discuss the variational and the Numerov methods used to
calculate the deuteron bound and $nn$ scattering wave functions.

\vspace*{1cm}
\subsection{Observables}
\label{subsec:obs}

The differential capture rate in the doublet initial hyperfine
state ${d\Gamma^D}/{dp}$ can be written as~\cite{Marcucci2011}
\begin{equation}
\frac{d\Gamma^D}{dp}=E_\nu^2\, \left[ 1-{E_\nu \over (m_\mu + m_d)}\right]
\,\frac{p^2d{\hat{\bf p}}}{8\pi^4}\,
\overline{|T_W|^2} \ ,
\label{eq:dgd}
\end{equation}
where ${\bf p}$ is the $nn$ relative momentum, and 
\begin{equation}
E_\nu=\frac{(m_\mu +m_d)^2 -4m_n^2-4 p^2}{2 (m_\mu +m_d)} \ ,
\label{eq:enu2}
\end{equation}
with $m_\mu$, $m_n$, and $m_d$ being the muon, neutron, and deuteron  masses.
The transition amplitude $\overline{|T_W|^2}$ reads~\cite{Marcucci2011}
\begin{equation}
\overline{|T_W|^2} = \frac{1}{2f+1}\sum_{s_1 s_2 h_\nu}\sum_{f_z}
|T_W(f,f_z;s_1,s_2, h_\nu)|^2 \ ,
\label{eq:hw2}
\end{equation}
where $f,f_z$ indicate the initial hyperfine state, fixed
here to be $f=1/2$, while $s_1$, $s_2$, and $h_\nu$ denote the
spin $z$-projection for the two neutrons and the neutrino helicity state.
In turn, $T_W(f,f_z;s_1,s_2, h_\nu)$ is given by
%
\begin{eqnarray}
T_W (f,f_z;s_1,s_2,h_\nu) &\equiv&
\langle nn, s_1, s_2; \nu, h_\nu \,|\, H_W \,|\,
(\mu,d);f,f_z \rangle
\nonumber \\
&\simeq& {G_V \over \sqrt{2}} \psi_{1s}^{\rm av}
\sum_{s_\mu s_d}
\langle {1 \over 2}s_{\mu}, 1 s_d | f f_z \rangle\,
l_\sigma(h_\nu,\,s_\mu)\,
\langle \Psi_{{\bf p}, s_1 s_2}(nn) | j^{\sigma}({\bf q}) |
\Psi_d(s_d)\rangle \ , \label{eq:h2ffz}
\end{eqnarray}
%
with $l_\sigma$ and $j^\sigma$ being the leptonic and
hadronic current densities, respectively~\cite{Marcucci2011},
written as
\begin{equation}
l_\sigma(h_\nu,\,s_\mu) \equiv
{\overline{u}}({\bf k}_\nu,h_\nu)\,\gamma_{\sigma}\, (1-\gamma_5)
u({\bf k}_\mu,s_{\mu}) \>\>\>,
\label{eq:lsigma}
\end{equation}
and
\begin{equation}
j^\sigma({\bf q})=\int {\rm d}{\bf x}\,
{\rm e}^{ {\rm i}{\bf q} \cdot {\bf x} }\,j^\sigma({\bf x})
\equiv (\rho({\bf q}),{\bf j}({\bf q}))
\label{eq:jvq} \>\>\>.
\end{equation}
Here the leptonic momentum transfer ${\bf q}$ is defined
as ${\bf q} = {\bf k}_\mu-{\bf k}_\nu \simeq -{\bf k}_\nu$.
Furthermore, $\Psi_d(s_d)$ and $\Psi_{{\bf p}, s_1 s_2}(nn)$ are the
initial deuteron and final $nn$ wave functions, respectively,
with $s_d$ indicating the deuteron spin $z$-projection.
Finally, in Eq.~(\ref{eq:h2ffz}), the function $\psi_{1s}^{\rm av}$
represents the $1s$ solution of the Schr\"odinger
equation for the initial muonic $\mu-d$ atom.
Since the muon is essentially at rest, it can be approximated
as~\cite{Marcucci2011,Walecka1995}
\begin{equation}
|\psi_{1s}^{\rm av}| \equiv\,  |\psi_{1s}(0)|\,=\,
\sqrt{{(\alpha\, \mu_{\mu d})^3\over \pi}} \ ,
\label{eq:psimud}
\end{equation}
where $\psi_{1s}(0)$ denotes the Bohr wave function
for a point charge $e$ evaluated at the origin, 
$\mu_{\mu d}$ is the reduced mass of the $(\mu,d)$ system,
and $\alpha=1/137.036$ is the fine-structure constant.

The final $nn$ wave fucntion can be expanded in partial waves as
%
\begin{equation}
\Psi_{{\bf p},s_1 s_2}(nn)=4\pi\sum_{S} \langle \frac{1}{2} s_1,\frac{1}{2}s_2 |
S S_z \rangle 
\sum_{L L_z J J_z}{\rm i}^L Y^*_{LL_z}({\hat{\bf p}}) 
\langle S S_z, L L_z | J J_z\rangle \,\overline{\Psi}_{nn}^{LSJJ_z}(p) \>\> ,
\label{eq:psinnpw}
\end{equation}
%
where $\overline{\Psi}_{nn}^{LSJJ_z}(p)$ is the $nn$ wave function with
orbital angular momentum $L L_z$, total spin $S S_z$, and total angular
momentum $J J_z$. In the present work, we restrict our study to the
$L=0$ state ($^1S_0$ in spectroscopic notation).

Using standard techniques as described in
Refs.~\cite{Marcucci2011,Walecka1995}, a multipole expansion
of the weak charge, $\rho({\bf q})$, and current,
${\bf j}({\bf q})$, operators can be performed, resulting in
%
\begin{eqnarray}
\langle \overline{\Psi}_{nn}^{LSJJ_z}(p) | \rho({\bf q}) | \Psi_d(s_d) \rangle 
&=&
\sqrt{4\pi}\sum_{\Lambda \geq 0}\sqrt{2\Lambda+1}\,\,{\rm i}^\Lambda
\frac{\langle 1 s_d, \Lambda 0 | J J_z\rangle}{\sqrt{2J+1}} C_\Lambda^{LSJ}(q) \ ,
\label{eq:c2} \\
\langle \overline{\Psi}_{nn}^{LSJJ_z}(p) | j_z({\bf q}) | \Psi_d(s_d) \rangle 
&=&
-\sqrt{4\pi}\sum_{\Lambda \geq 0}\sqrt{2\Lambda+1}\,\,{\rm i}^\Lambda
\frac{\langle 1 s_d, \Lambda 0 | J J_z\rangle}{\sqrt{2J+1}} L_\Lambda^{LSJ}(q) \ ,
\label{eq:l2} \\
\langle \overline{\Psi}_{nn}^{LSJJ_z}(p) | j_\lambda({\bf q}) | 
\Psi_d(s_d) \rangle 
&=&
\sqrt{2\pi}\sum_{\Lambda \geq 1}\sqrt{2\Lambda+1}\,\,{\rm i}^\Lambda
\frac{\langle 1 s_d, \Lambda -\lambda | J J_z\rangle}{\sqrt{2J+1}} \nonumber \\
&\times&
[-\lambda M_\lambda^{LSJ}(q)+E_\Lambda^{LSJ}(q)]\ ,
\label{eq:em}
\end{eqnarray}
%
where $\lambda=\pm 1$, and $C_{\Lambda}^{LSJ}(q)$, $L_{\Lambda}^{LSJ}(q)$,
$E_{\Lambda}^{LSJ}(q)$ and $M_{\Lambda}^{LSJ}(q)$ denote the reduced matrix 
elements (RMEs) of the Coulomb ($C$), longitudinal ($L$), transverse
electric ($E$) and transverse magnetic ($M$) multipole operators, as defined
in Ref.~\cite{Marcucci2011}. Since the weak charge and current
operators have scalar/polar-vector $(V)$ and pseudo-scalar/axial-vector $(A)$
components, each multipole consists of the sum of $V$ and $A$ terms,
having opposite parity under space inversions. Given that in this study
only the $^1S_0$ contribution is considered, the only contributing
multipoles are $C_1(A)$, $L_1(A)$, $E_1(A)$, $M_1(V)$, where the
superscripts $LSJ$ have been dropped.

In order to calculate the differential capture rate 
$d\Gamma^D/dp$ in Eq.~(\ref{eq:dgd}), we need to integrate
over ${\hat{\bf p}}$. This is done numerically
using Gauss-Legendre of the order of 10, so that an accuracy to better
than 1 part in $10^3$ can be achieved.
Finally, the total capture rate $\Gamma^D$ is obtained as
\begin{equation}
  \Gamma^D=\int_{0}^{p_{max}}\frac{d\Gamma^D}{dp}dp \ ,
\end{equation}
where $p_{max}$ is the maximum value of the momentum $p$. In order to find
the smallest needed number of grid points to reach convergence, we have
computed the capture rate by integrating over several grids starting from a
minimum value of 20 points up to a maximum of 80. We have verified that
the results obtained integrating over 20 or 40 points differ of about
0.1 s$^{-1}$, while the ones obtained with 40, 60 and 80 points
differ by less than 0.01 s$^{-1}$. Therefore, we have used 60
grid points in all the studied cases mentioned below.

\hfill

\subsection{Nuclear potentials and currents}
\label{subsec:pot-curr}

In this study we consider four different nuclear interaction models,
and consistent weak current operators, derived in $\chi$EFT.
We decided to concentrate our attention on the recent
local $r$-space potentials of Ref.~\cite{Piarulli2016}
(see also Ref.~\cite{Piarulli2020} for a recent review). The motivation
behind this choice is mostly related to the fact that in the future we plan
to use this same formalism to the $pp$ reaction,
for which the Coulomb
interaction, and also electromagnetic higher order contributions,
play a significant role at the accuracy level reached by theory.
The possibility to work in $r$-space is clearly
an advantage compared with momentum-space, which would be the unavoidable
choice when using non-local potentials. However, in momentum-space the full
electromagnetic interaction between the two protons is not easy to
be taken into account.
The potentials of Ref.~\cite{Piarulli2016}, which we will refer to as
Norfolk potentials (denoted as NV), are chiral interactions that include,
beyond pions and nucleons, also $\Delta$-isobar degrees of freedom explicitly.
The short-range (contact) part of the interaction receives contributions
at leading order (LO), next-to-leading order (NLO) and
next-to-next-to-next-to-leading order (N3LO), while the long-range
components arise from one- and two-pion exchanges,
and are retained
up to next-to-next-to-leading order (N2LO). 
By truncating the expansion at N3LO, there are 26 LECs
which have been fitted to the $NN$
Granada database~\cite{Navarro2013,Navarro2014a,Navarro2014b},
obtaining two classes of Norfolk potentials, depending on the range
of laboratory energies over which the fits have been carried out:
the NVI potentials have been fitted in the range 0--125 MeV, while for the NVII
potentials the range has been extended up to 200 MeV. 
For each class of potential, two cutoff functions
$C_{R_S}(r)$ and $C_{R_L}(r)$ have been used to regularize the
short- and long-range components, respectively. These functions
have been defined as
\begin{eqnarray}
  C_{R_S}(r)&=&\frac{1}{\pi^{\frac{3}{2}}R_S^3}{\rm e}^{-(r/R_S)^2}
  \ , \label{eq:CRS} \\
  C_{R_L}(r)&=&1-\frac{1}{(r/R_L)^6{\rm e}^{(r-R_L)/a_L}+1}
  \ , \label{eq:CRL}
\end{eqnarray}
with $a_L\equiv R_L/2$. Two different sets of cutoff values
have been considered,
$(R_S;R_L)=(0.7;1.0)$ and $(0.8;1.2)$, and the resulting models have been
labelled ``a'' and ``b'', respectively. All these potentials are very
accurate: in fact, the $\chi^2$/datum for the 
NVIa, NVIIa, NVIb, and NVIIb potentials are, respectively, 1.05, 1.37, 1.07,
and 1.37~\cite{Piarulli2016}.

We turn now our attention to the weak transition operators.
When only the $^1S_0$ $nn$ partial wave is included, we have seen that 
the contributing multipoles are
$C_1(A)$, $L_1(A)$, $E_1(A)$ and $M_1(V)$.
Consequently, the weak vector charge operator is of no interest
in the process under consideration, and we will not discuss it here.
The weak vector current entering $M_1(V)$ can be obtained from the isovector
electromagnetic current, performing a rotation in the isospin space, i.e.\
with the substitutions
\begin{eqnarray}
  \tau_{i,z} &\Rightarrow & \tau_{i,\pm}=(\tau_{i,x}\pm i \tau_{i,y})/2 \ ,
  \label{eq:tauipm} \\
  ({\bm \tau}_i\times{\bm\tau}_j)_z&\Rightarrow&
  ({\bm \tau}_i\times{\bm\tau}_j)_\pm =({\bm \tau}_i\times{\bm\tau}_j)_x
  \pm i ({\bm \tau}_i\times{\bm\tau}_j)_y \label{eq:tauvectortau}\ .
\end{eqnarray}
Therefore, we will review the various contributions
to the electromagnetic current, even if, in fact,
we are interest only to their isovector components.
The electromagnetic current operators up
to one loop have been most recently reviewed in Ref.~\cite{Gnech2022}. Here
we only give a synthetic summary. Following the notation
of Ref.~\cite{Gnech2022}, we denote with $Q$ the generic low-momentum scale.
The LO contribution, at order $Q^{-2}$, consists of the
single-nucleon current, while at the NLO, or at order $Q^{-1}$, there
is the one-pion-exchange (OPE) contribution.
The relativistic
correction to the LO single-nucleon current provides the
first contribution of order $Q^0$ (N2LO). Furthermore,
since the Norfolk interaction models retain explicitly $\Delta$-isobar
degrees of freedom, we take into account also the N2LO currents
originating from explicit $\Delta$ intermediate states.
Finally, the currents at order $Q^1$ (N3LO)
consist of (i) terms generated by minimal substitution in the four-nucleon
contact interactions involving two gradients of the nucleon fields
and by non-minimal couplings to the electromagnetic field; (ii)
OPE terms induced by $\gamma\pi N$ interactions of sub-leading order; and
(iii) one-loop two-pion-exchange terms.
A thourough discussion of all these contribuions as well as 
their explicit expressions
can be found in Ref.~\cite{Gnech2022}. Here we only remark that
(i) the various contributions are derived in momentum space and
have power law behavior at large momenta, or short range. Therefore,
they need to be
regularized. The procedure adopted here, as in Ref.~\cite{Gnech2022}, is
to carry out first the Fourier transforms of the various terms. This results
in $r$-space operators which are highly singular 
at vanishing inter-nucleon separations. Then the singular behavior
is removed by multiplying the various terms by appropriate
$r$-space cutoff functions, identical to those of the
Norfolk potentials of Ref.~\cite{Piarulli2016}.
More details can be found in Refs.~\cite{Schiavilla2019,Gnech2022}.
(ii) There are 5 LECs in the electromagnetic currents which do not
enter in the nuclear potentials and need to be fitted
using electromagnetic observables. These LECs enter the current
operators at N3LO, in particular two of them are present in the
currents arising from non-minimal couplings to the electromagnetic field,
and three of them are present in the sub-leading isoscalar and isovector
OPE contributions. In this study, these LECs are determined by a
simultaneous fit to the $A= 2–3$ nuclei magnetic moments and to the
deuteron threshold electrodisintegration
at backward angles over a wide range of momentum transfers~\cite{Gnech2022}.
In this work we used the LECs labelled with set A in Ref.~\cite{Gnech2022}.

The axial current operators used in the present work are the ones of
Ref.~\cite{Baroni2018}. They include the LO term, of order $Q^{-3}$, which
arises from the single-nucleon axial current, and the N2LO and N3LO terms
(scaling as $Q^{-1}$ and $Q^{0}$, respectively), consisting of the
relativistic corrections and $\Delta$ contributions at N2LO, and of OPE
and contact-terms at N3LO. Note that at NLO, here of
order $Q^{-2}$, there is no contribution in $\chi$EFT. The explicit
$r$-space expression of these operators can be found in Ref.~\cite{Baroni2018}.
Here we only remark that all contributions have been regularized at
short and long range consistently with the regulator functions
used in the Norfolk potentials. Furthermore,
the N3LO contact-term presents a LEC, here
denoted with $z_0$ (but essentially equal to the
$d_R$ LEC mentioned in Sec.~\ref{sec:intro}), defined as
\begin{equation}
z_0=\frac{g_A}{2}\frac{m_{\pi}^2}{f_{\pi}^2}\frac{1}{(m_{\pi}R_S)^3}\left[-\frac{m_{\pi}}{4g_A\Lambda_{\chi}}c_D+\frac{m_{\pi}}{3}\left(c_3+2c_4\right)+\frac{m_{\pi}}{6m}\right]\ .\label{z0}
\end{equation}
Here $g_A=1.2723(23)$ is the single-nucleon axial coupling constant, $m=938.9$ MeV the nucleon
mass, $m_\pi=138.04$ MeV and $f_\pi=97.4$ MeV
the pion mass and decay constant, $\Lambda_\chi\sim 1$ GeV the chiral-symmetry
breaking scale, and $c_3=-0.79$ and $c_4=1.33$ two LECs entering the $\pi\pi N$ Lagrangian
at N2LO  and taken from the fit of the  pion-nucleon scattering data with $\Delta$-isobar as
explicit degrees of freedom~\cite{Krebs2007}.
As mentioned above, $c_D$ is one of the two LECs which enter the three-nucleon
interaction, the other being denoted with $c_E$. The two LECs
$c_D$ (and consequently $z_0$) and $c_E$ have been fitted
to simultaneously reproduce the experimental trinucleon binding
energies and the central value of the Gamow-Teller matrix element
in triton $\beta$-decay.
The explicit values for $c_D$ are 
$-0.635$, $-4.71$, $-0.61$ and $-5.25$ for the NVIa, NVIb, NVIIa, and NVIIb
potentials respectively.

The nuclear axial charge has a much simpler structure
compared to the axial and vector currents, and we have used
the operators as derived in Ref.~\cite{Baroni2016}.
At LO, i.e.\ at order $Q^{-2}$, it retains the one-body term, which
gives the most important contribution. At NLO (order $Q^{-1}$) the OPE
contribution appears, which however has been found almost
negligible in this study. The N2LO 
contributions (order $Q^{0}$) exactly vanish,
and at N3LO (order $Q^1$) there are two-pion exchange terms and new contact terms where new LECs appear. The  N3LO has not been included in the calculation, since the new LECs have not been fixed yet.
However, we have found the contribution of $C_1(A)$ to be two orders of magnitude smaller compared 
to the one from the other multipoles. Therefore, the effect of the axial current correction at N3LO can be safely disregarded. 

All the axial charge and current contributions are multiplied
by the single-nucleon axial coupling constant, $g_A(q_\sigma^2)$, written
as function of the squared of the four-momentum transfer $q_\sigma^2$.
Contrary to the
triton $\beta$-decay, in the case of the muon capture
on deuteron, the four-momentum transfer is quite large. The
dependence of $g_A(q_\sigma^2)$ on $q_\sigma^2$ is therefore crucial and,
as already mentioned in Sec.~\ref{sec:intro}, it is
a source of theoretical uncertainty in this study.
In the past, it has been used for $g_A(q_\sigma^2)$
a dipole form~\cite{Marcucci2011},
but in Ref.~\cite{Meyer2016}
it has been argued that the dipole form introduces an uncontrolled systematic
error in estimating the value of the axial form
factor. Alternatively, it has been proposed to use
the small-momenta expansion, which leads to the expression
of Eq.~(\ref{esp}).
We have decided to use in our study
the new parameterization for $g_A(q_{\sigma}^2)$
of Eq.~(\ref{esp}),
but with a slightly smaller uncertainty on the 
the axial charge radius $r_A$ compared with Ref.~\cite{Meyer2016},
as discussed in Ref.~\cite{Hill2018}. In this work, $r_A$ has been chosen
as the weighted average of the values obtained by two independent procedures
having approximately the same accuracy, about $50 \%$. One procedure
is the one of Ref.~\cite{Meyer2016}, and uses for
the axial form factor a convergent expansion given
by
\begin{equation}
g_A(q_{\sigma}^2)=\sum_{k=0}^{k_{max}}a_kz(q_{\sigma}^2)^k\label{gaz}\ ,
\end{equation}
where the variable $z(q_{\sigma}^2)$ is defined as
\begin{equation}
  z(q_{\sigma}^2)=\frac{\sqrt{t_{cut}-q_{\sigma}^2}-\sqrt{t_{cut}-t_0}}
  {\sqrt{t_{cut}-q_{\sigma}^2}+\sqrt{t_{cut}-t_0}}\ ,
  \label{eq:zq}
\end{equation}
with $t_{cut}=9\;m_{\pi}^2$ and $-\infty<t_0<t_{cut}$. In Eq.~(\ref{gaz}),
$a_k$ are the expansion parameters which encode the nuclear structure
information and need to be experimentally fixed. From $g_A(q_{\sigma}^2)$ in
Eq.~(\ref{gaz}), we can obtain $r_A^2$ as~\cite{Meyer2016}
\begin{equation}
  \frac{1}{6}r_A^2\equiv\frac{1}{g_A(0)}\left.\frac{dg_A(q_\sigma^2)}{dq_{\sigma}^2}\right|_{q_{\sigma}^2=0} \ .
  \label{eq:ra_dg}
\end{equation}
The value for $r^2_A$ is obtained fitting experimental data
of neutrino scattering on deuterium and it is found to be
$r^2_A\left(z\right.$ exp.$\left.\nu\right)=0.46(22)$
fm$^2$~\cite{Meyer2016}. 

Alternatively it is possible to obtain $r^2_A$ from experiments on
muonic capture on proton, as done by the MuCap Collaboration.
To date these experiments are characterized
by an overall accuracy of 1\%, but a future experiment plans to reduce
this uncertainty to about 0.33\%~\cite{Hill2018}. In this case,
$r^2_A\left(\right.$MuCap$\left.\right)=0.46(24)$ fm$^2$~\cite{Hill2018}.
In order to take into account both
$r^2_A\left(z\right.$ exp.$\left.\nu\right)$ and
$r^2_A\left(\right.$MuCap$\left.\right)$, we adopted for $r^2_A$ the
value $r_A^2=0.46(16)$ fm$^2$, as suggested in Ref.~\cite{Hill2018}.
The uncertainty on $r_A^2$ remains quite large, of about 35$\%$,
but it is slightly smaller than the one of Ref.~\cite{Meyer2016},
which has been adopted in the study of Ref.~\cite{Acharya2018}. The
consequences on the error budget will be discussed in Sec.~\ref{sec:results}.

\vspace*{0.5cm}
\subsection{Nuclear wave functions}
\label{subsec:nuclwf}

The calculation of the nuclear wave functions of the deuteron and
$nn$ systems have been first of all performed using the variational method
described in Ref.~\cite{Marcucci2011}, where all the details of the
calculation can be found. Here we summarize only the main steps.

The deuteron wave function can be written as

\begin{equation}
  \Psi_d(\mathbf{r},j_z)=\sum_{\alpha}\sum_{i=0}^{M-1}c_{\alpha,i}\,f_i(r)\,
  \mathcal{Y}_{\alpha}(\hat{\mathbf{r}}),\label{psi}
\end{equation}
where the channels $\alpha\equiv (l; s; J; t)$ denotes the
deuteron quantum numbers, with the combination $(l=0,2; s=1; J=1; t=0)$
corresponding to $\alpha=1,2$, respectively, and the functions
$\mathcal{Y}_{\alpha}(\hat{\mathbf{r}})$ are given by
\begin{equation}
  \mathcal{Y}_{\alpha}(\hat{\mathbf{r}})\equiv \left[Y_{l}(\hat{\mathbf{r}})
    \otimes\chi_{s}\right]_{J J_z}\xi_{t t_z}\ . \label{armsfvet}
\end{equation}
The $M$ radial functions $f_{i}(r)$,
normalized to unity, with $i=0,\cdots,M-1$, are written as
\begin{equation}
  f_i(r)=\sqrt{\frac{i!\gamma^3}{\left(i+2\right)!}}
  e^{-\frac{\gamma}{2}r\;(2)\!}L_i(\gamma r)
  \label{uf} \ ,
\end{equation}
where $\gamma$ is a non-variational parameter chosen to
be~\cite{Marcucci2011} $\gamma=0.25\;\rm{fm}^{-1}$ and
$^{(2)}\!L_i(\gamma r)$ are the Laguerre polynomials of the second
type~\cite{abramowitz1964handbook}.
The unknown coefficients $c_{\alpha,i}$ are obtained
using the Rayleigh-Ritz variational principle, i.e.\ imposing the
condition
\begin{equation}
\frac{\partial}{\partial c_{\alpha,i}}\langle\Psi_d|H+B_d|\Psi_d\rangle=0 \ ,
\end{equation}
where $H$ is the Hamiltonian and $B_d$ is the deuteron binding energy. This
reduces to an eigenvalue-eigenvector problem, which can be solved
with standard numerical techniques~\cite{Marcucci2011}.

The $nn$ wave function $\overline{\Psi}_{nn}^{LSJJ_z}(p)$ in
Eq.~(\ref{eq:psinnpw}) is written as a sum of a core wave function
$\Psi^{c}(p)$, and of an asymptotic wave function $\Psi^{a}(p)$,
where we have dropped the superscript $LSJJ_z$
for ease of presentation. The core wave function $\Psi^{c}(p)$
describes the $nn$
scattering state where the two nucleons are close to each other, and
is expanded on a basis of Laguerre polynomials, similarly to what we have done
for the deuteron wave function. Therefore
\begin{equation}
  \Psi^{c}(p)=\sum_{i=0}^{M-1}d_{i}(p)f_i(r)
  \mathcal{Y}_{\alpha}(\hat{\mathbf{r}})\ ,\label{psicore}
\end{equation}
where $f_i(r)$ and $\mathcal{Y}_{\alpha}(\hat{\mathbf{r}})$ are defined in
Eqs.~(\ref{uf}) and (\ref{armsfvet}), respectively. Note
that $\alpha\equiv L=0; S=0, J=0, J_z=0$. In the unknown
coefficients $d_{i}(p)$ we have kept explicitly the dependence on $p$.

The asymptotic wave function $\Psi^{a}(p)$
describes the $nn$ scattering system in the asymptotic region, where the
nuclear potential is negligible. Consequently, it can be written
as a linear combination of regular (Bessel) and irregular (Neumann) spherical
functions, denoted as $j_L(pr)$ and $n_L(pr)$, respectively, i.e.\
\begin{equation}
  \Psi^{a}(p)=
  \tilde{F}_{L}(pr)\mathcal{Y}_{\alpha}(\hat{\mathbf{r}})+
  \sum_{L'}R_{LL'}\;\tilde{G}_{L'}(pr)
  \mathcal{Y}_{\alpha'}(\hat{\mathbf{r}})\ ,\label{psiasymptotic}
\end{equation}
where $R_{LL'}$ is the reactance matrix, and 
$\tilde{F}_{L'}(pr)$ and $\tilde{G}_{L'}(pr)$ are defined as 
\begin{eqnarray}
\tilde{F}_{L'}(pr)&\equiv&\frac{j_L(pr)}{p^L}\ , \label{besselregular}\\
\tilde{G}_{L'}(pr)&\equiv&n_L(pr)(1-e^{-\epsilon r})^{2L+1}p^{L+1}\ ,
\label{besselirregular}
\end{eqnarray}
so that they are well defined for $p\rightarrow0$ and $r\rightarrow0$.
The function $(1-e^{-\epsilon r})^{2L+1}$ has been found to be an appropriate
regularization factor at the origin for $n_L(pr)$. We use the value
$\epsilon=0.25\;\rm{fm}^{-1}$ as in Ref.~\cite{Marcucci2011}.
To be noticed that since here $L=L'=0$ the reactance matrix is in fact
just a number, and $R_{00}=\tan\delta_0$, $\delta_0$ being
the phase shift.

In order to determine 
the coefficients $d_{i}(p)$ in Eq.~(\ref{psicore}) and the reactance matrix
$R_{LL'}$ in Eq.~(\ref{psiasymptotic}), we use the Kohn variational
principle~\cite{Kohn1948}, which states that the functional
\begin{equation}
  \left[R_{LL'}(p)\right]=R_{LL'}(p)-\frac{m_n}{\hbar^2}
  \langle\overline{\Psi}_{\alpha'}(p)|H-E|\overline{\Psi}_{\alpha}(p)\rangle\ ,
  \label{Kohn}
\end{equation}
is stationary with respect to $d_{i}(p)$ and $R_{LL'}$.
In Eq.~(\ref{Kohn}) $E$ is the $nn$ relative energy
($E = p^2/m_n$, $m_n$ being the neutron mass)
and $H$ is the Hamiltonian operator.
Performing the variation, a system of linear inhomogeneous equations for
$d_{i}(p)$  and a set of algebraic equations for $R_{LL'}$
are derived. These equations are solved by standard techniques. 
The variational results presented in the following section have been 
    are obtained using $M=35$ for both the deuteron and the $nn$ scattering wave
    functions.

In order to test the validity of the variational method
and its numerical accuracy,
in this work we have used also the Numerov method both for the deuteron
and the $nn$ wave functions. 

For the deuteron wave function, we have used the so called renormalized
Numerov method, based on the work of Ref.~\cite{Johnson1978}.
Within this method, the Schr\"odinger equation is rewritten as
\begin{equation}
\left[I\frac{d^2}{dx^2}+Q(x)\right]\Psi(x)=0\label{Sequation}\ ,
\end{equation}
where $I$ is the identity matrix, $Q(x)$ is a matrix defined as
\begin{equation}
  Q(x)=\left(\frac{2\mu}{\hbar^2}\right)\left[EI-V(x)\right]
  \label{qmatrix}\ ,
\end{equation}
and $\Psi(x)$ is also a matrix whose columns are the independent solutions
of the Schr\"odinger equation with non assigned boundary conditions on the
derivatives.
In Eq.~(\ref{qmatrix}), $\mu$ is the $np$ reduced mass,
$E\equiv -B_d$,
and $V(x)$ is the sum of the $np$ nuclear potential $V^{np}(x)$ and the
centrifugal barrier, i.e.\
\begin{equation}
V(x)=V^{np}(x)+\frac{\hbar^2l(l+1)}{2\mu r^2}\ .
\end{equation}
The Schr\"odinger equation is evaluated on a finite and discrete
grid with constant step $h$. The boundary conditions require to know
the wave function at the initial and final grid points, given by
$x_0=0$ and $x_N=Nh$ respectively. Specifically,
it is assumed that $\Psi(0)=0$ and $\Psi(Nh)=0$.
No condition on first derivatives are imposed. 

Eq.~(\ref{Sequation}) can be rewritten equivalently as~\cite{Johnson1978}
\begin{equation}
  \left[I-T(x_{n+1})\right]\Psi(x_{n+1})-\left[2I+10T(x_{n})\right]\Psi(x_{n})+
  \left[I-T(x_{n-1})\right]\Psi(x_{n-1})=0\label{eq:numerov}\ ,
\end{equation}
where $x_n\in A$, $A\equiv\left(x_0,x_N\right)$, and $T(x_n)$ is a
$2\times2$ matrix defined as~\cite{Johnson1978}
\begin{equation}
T(x_n)=-\frac{h^2}{12}Q(x_n)\ .
\label{txn}
\end{equation}  
To be noticed that Eq.~(\ref{eq:numerov}) is in fact the natural extension
to a matrix formulation of the ordinary Numerov algorithm (see Eq.~(\ref{3p}) below).

By introducing the matrix
$F(x_n)$ as~\cite{Johnson1978}
\begin{equation}
F(x_n)=[I-T(x_n)]\Psi(x_n)\label{F}\ ,
\end{equation}
Eq.~(\ref{eq:numerov}) can be rewritten as
\begin{equation}
F(x_{n+1})-U(x_n)F(x_n)+F(x_{n-1})=0\label{recorsive}\ ,
\end{equation}
where the matrix $U(x_n)$ is given by
\begin{equation}
U(x_n)=\left[I-T(x_n)\right]^{-1}\left[2I+10\;T(x_n)\right]\ .
\label{uxn}
\end{equation}
Furthermore, we introduce the matrices 
$R(x_n)$ and $\hat{R}(x_n)$, 
defined as~\cite{Johnson1978}
\begin{eqnarray}
R(x_n)&=&F(x_{n+1})F^{-1}(x_n)\label{R}\ ,\\
\hat{R}(x_n)&=&F(x_{n-1})F^{-1}(x_n)\label{Rcap}\ ,
\end{eqnarray}
and their inverse matrices as
\begin{eqnarray}
R^{-1}(x_n)&=&F(x_n)F^{-1}(x_{n+1})\ ,\label{R0}\\
\hat{R}^{-1}(x_n)&=&F(x_n)F^{-1}(x_{n-1})\label{Rcap0}\ .
\end{eqnarray} 
By using the definitions~(\ref{R}) and~(\ref{Rcap}),
it is possible to derive from 
Eq.~(\ref{recorsive})
the following recursive relations
\begin{eqnarray}
R(x_n)&=&U(x_n)-R^{-1}(x_{n-1})\label{R1}\ ,\\
\hat{R}(x_n)&=&U(x_n)-\hat{R}^{-1}(x_{n+1})\label{Rcap1}\ .
\end{eqnarray}
We now notice that, since $\Psi(0)=0$,
Eq.~(\ref{F}) implies that $F(0)=0$ and, consequently,
from Eq.~(\ref{R0}) it follows that $R^{-1}(0)=0.$ Similarly,
since $\Psi(Nh)=0$, from Eqs.~(\ref{F}) and (\ref{Rcap0}) we obtain
that $\hat{R}^{-1}(Nh)=0$. Starting from the $R^{-1}(0)$ and $\hat{R}^{-1}(Nh)$
values, and iteratively using Eqs.~(\ref{R1}) and (\ref{Rcap1}),
it is possible to calculate the 
$R(x_m)$ and $\hat{R}^{-1}(x_{m+1})$ values up to a matching
point $x_m$, so that 
the interval $A$ remains divided into two sub-intervals, 
$A_1\equiv\left[x_0,x_{m+1}\right]$ and $A_2\equiv\left[x_m,x_N\right]$.
These values are needed in order to calculate the
deuteron binding energy and its wave function. In fact, 
assuming we knew the deuteron binding energy $B_d\equiv -E$
for a given potential,
then we could integrate Eq.~(\ref{Sequation}) in the two sub-intervals
$A_1$ and $A_2$, obtaining the outgoing (left)
solution $\Psi_l(x_n)$ in $A_1$, and the incoming (right)
solution $\Psi_r(x_n)$ in $A_2$. If $B_d$ were a true
eigenvalue, then the function 
$\Psi(x_n)$ and its derivative have to be continuous in $x_m$.
The wave function continuity at two consecutive points, for example $x_m$ and
$x_{m+1}$, implies that
\begin{eqnarray}
  \Psi_l(x_m)\cdot\mathbf{l}&=&\Psi_r(x_m)\cdot\textbf{r}\equiv\psi(x_m)\ ,
  \label{xm}\\
  \Psi_l(x_{m+1})\cdot\mathbf{l}&=&\Psi_r(x_{m+1})\cdot\textbf{r}
  \equiv\psi(x_{m+1})\ ,
  \label{xm1}
\end{eqnarray}
where $\mathbf{l}$ and $\mathbf{r}$ are two unknown vectors.
Multiplying Eq.~(\ref{xm1}) by $\left[I-T(x_{m+1})\right]$ and
using Eq.~(\ref{F}), we obtain
%
%
\begin{equation}
  F_l(x_{m+1})\cdot\mathbf{l}=
  F_r(x_{m+1})\cdot\textbf{r}\equiv f(x_{m+1})\label{xm12}\ .
\end{equation}
Similarly, from Eq.~(\ref{xm}) we can write
\begin{equation}
F_l(x_m)\cdot\mathbf{l}=F_r(x_m)\cdot\textbf{r}\equiv f(x_m)\ . \label{fxm}
\end{equation} 
Using Eq.~(\ref{R}) with $x_n=x_{m}$, for the outgoing solution, and
Eq.~(\ref{Rcap}) with $x_n=x_{m+1}$, for the incoming solution, we can write
\begin{eqnarray}
F_l(x_{m+1})&=&R(x_m)F_l(x_m)\ ,\label{f1}\\
F_r(x_{m+1})&=&\hat{R}^{-1}(x_{m+1})F_r(x_m)\ .\label{f2}
\end{eqnarray}
By replacing Eqs.~(\ref{f1}) and (\ref{f2}) into Eq.~(\ref{xm12})
and
%
using Eq.~(\ref{fxm}),
we obtain that
\begin{equation}
  R(x_m)f(x_m)=\hat{R}^{-1}(x_{m+1})f(x_m)\ ,
  \label{rxmf1}
\end{equation}
or equivalently that
\begin{equation}
\left[R(x_m)-\hat{R}^{-1}(x_{m+1})\right]f(x_m)=0\label{detrr-1}\ .
\end{equation}
Non-trivial solution is only admitted if the above equation satisfies
the following condition
\begin{equation}
  det\left[R(x_m)-\hat{R}^{-1}(x_{m+1})\right]=0 \ .
  \label{determinant}
\end{equation}
This determinant is a function of the energy $E$, i.e.
\begin{equation}
  det(E)=det\left[R(x_m)-\hat{R}^{-1}(x_{m+1})\right]\ .
  \label{determE}
\end{equation}
Therefore, 
we proceed as follows: starting from an initial trial value $E_1$, we
calculate $det(E_1)$. Fixed a tolerance factor $\epsilon$, for example
$\epsilon=10^{-16}$, if $det(E_1)\leq\epsilon$ we assume $E_1$ being
the eigenvalue, otherwise we compute the determinant for a second energy
value $E_2$.
If $det(E_2)\leq\epsilon$, we take the deuteron binding energy as
$B_d=-E_2$, otherwise it is necessary 
to repeat the procedure iteratively until $det(E_i)\leq\epsilon$.
For the iterations after the second one, the energy is chosen through the
relation
\begin{equation}
  E_i=E_{i-2}-det(E_{i-1})\frac{E_{i-2}-E_{i-1}}{det(E_{i-2})-det(E_{i-1})}\ ,
  \label{ei}
\end{equation}
which follows from a linear interpolation procedure.
The procedure stops when $det(E_i)\leq\epsilon$, and the deuteron
binding energy is taken to be $B_d=-E_i$.

In order to calculated the $S$- and $D$-wave components of the
reduced radial wave function, denoted as $u_0(x_n)$ and $u_2(x_n)$
respectively, we notice that they are the two components of the vector
$\psi(x_n)$, defined in Eq.~(\ref{xm}) at the point $x_m$.
The starting point is to assign an arbitrary value to one of the two
components of the vector function $f(x_m)$ (see Eq.~(\ref{fxm})).
Since $R(x_m)$ and $\hat{R}^{-1}(x_{m+1})$ are known, the value of the other
component is fixed by Eq.~(\ref{detrr-1}). 
By defining the outgoing function as $f(x_n)=F(x_n)\cdot\mathbf{l}$,
from Eq.~(\ref{R}) it follows that
\begin{equation}
f(x_n)=R^{-1}(x_n)f(x_{n+1})\ ,\label{fo}
\end{equation}
where $n=m-1,...,0$. 
Similarly we can proceed for the incoming function. By defining it as
$f(x_n)=F(x_n)\cdot\mathbf{r}$, from Eq.~(\ref{Rcap}) we have that
\begin{equation}
f(x_n)=\hat{R}^{-1}(x_n)f(x_{n-1})\ ,\label{fi}
\end{equation}
where $n=m+1,...,N$. 
At this point, the vector function $f(x_n)$
can be calculated $\forall\;x_n\in\left[x_0,x_N\right]$, through
Eqs.~(\ref{fo}) and (\ref{fi}).
The $u_0(x_n)$ and $u_2(x_n)$ functions are given from $f(x_n)$ by
\begin{equation}
  \psi(x_n)=\left[I-T(x_n)\right]^{-1}f(x_n)\ .
  \label{psixn}
\end{equation}
Finally, the deuteron wave function is normalized to unity.

The single-channel Numerov method, also known as
a three-point algorithm, has been used to calculate the $nn$
wave function. Although the method is quite well known, in order
to provide a comprehensive review of all the approaches
to the $A=2$ systems, we briefly summarize its main steps.
Again, we start by defining a finite and discrete interval $I$,
with constant step $h$, characterized by the initial and final points,
$x_0=0$ and $x_N=N h$.
Then,
the Schr\"odinger equation can be cast in the form
\begin{equation}
  u''(x_n)\equiv \left.\frac{d^2\;u(x)}{dx^2}\right|_{x=x_n}=W(x_n)u(x_n)
  \label{S}\ ,
\end{equation}
where
\begin{equation}
  W(x_n)=\left(\frac{2\mu}{\hbar^2}\right) V(x_n)-p^2\ ,
  \label{w}
\end{equation}
being $V(x_n)$ the nuclear potential and $p$ the $nn$ relative momentum.
In order to solve Eq.~(\ref{S}), it is convenient to introduce
the function $z(x_n)$, defined as
\begin{equation}
z(x_n)=u(x_n)-\frac{h^2}{12}u''(x_n)\ .\label{z}
\end{equation}
By replacing Eq.~(\ref{S}) into Eq.~(\ref{z}), $z(x_n)$ can be rewritten as
\begin{equation}
z(x_n)=\left(1-\frac{h^2}{12}W(x_n)\right)u(x_n)\ .\label{z1}
\end{equation}
By expanding $z(x_{n-1})$ and $z(x_{n+1})$ in an interval around the point
$x_n$ in a Taylor series up to $O(h^4)$, 
%
and adding together the two expressions, we obtain 
%
%
\begin{equation}
z(x_{n+1})=2z(x_n)-z(x_{n-1})+h^2u''(x_n)+O(h^6)\label{3p}\ .
\end{equation}
This is a three-point relation: once the $z(x_{n-1})$ and $z(x_n)$
values are known, after calculating $u''(x_n)$ using Eq.~(\ref{S}),
we can compute $z(x_{n+1})$ at the order $O(h^6)$. 

By fixing the values
$u(0)=0$ and $u(h)=h$, we consequently know $z(0)$ and $z(h)$, i.e.\
\begin{eqnarray}
u(0)=0&\Longrightarrow &z(0)=0\label{uz0} \ , \\
u(h)=h&\Longrightarrow &z(h)=\left(1-\frac{h^2}{12}W(h)\right)u(h)\label{uz1} \ ,
\end{eqnarray}
and $u''(h)$ is obtained by Eq.~(\ref{S}). Then,
$z(2h)$ is obtained from Eq.~(\ref{3p}), and consequently
\begin{equation}
u(2h)=\frac{z(2h)}{\left[1-(h^2/12)W(2h)\right]}\label{uricorsivo}\ ,
\end{equation}
where $W(2h)$ is given by Eq.~(\ref{w}).
Eq.~(\ref{uricorsivo}) can be used again to determine the $u(3h)$ value,
and, proceeding iteratively, the $S$-wave scattering reduced
radial wave function is fully determined except for an overall normalization
factor. This means that
for a sufficiently large value of $x_n\in A$, denoted as $x_{\overline{n}}$,
we can write
\begin{equation}
  u(x_{\overline{n}})=N\left[j_0(kx_{\overline{n}})+
    \tan{\delta_0}\,n_0(kx_{\overline{n}})\right]\label{tan}\ .
\end{equation}
where $N$ is the sought normalization constant, and the phase shift $\delta_0$
can be computed taking the ratio between Eq.~(\ref{tan}) written for
$x_{\overline{n}}$ and the same equation written for $x_m$, $m$ being close to $\overline{n}$, so that
\begin{eqnarray}
  \tan{\delta_0}=\frac{u(x_m)j_0(kx_{\overline{n}})-u(x_{\overline{n}})
    j_0(kx_{m})}{u(x_{\overline{n}})n_0(kx_{m})-u(x_m)n_0(kx_{\overline{n}})}\ .
\end{eqnarray}
Finally, using Eq.~(\ref{tan}), the normalization constant $N$ 
is given by
\begin{equation}
  N=u(x_{\overline{n}})/\left[
    j_0(kx_{\overline{n}})+\tan{\delta_0}n_0(kx_{\overline{n}})\right]\ ,
\end{equation}
so that the function $u(x_n)$ turns out to be normalized to unitary flux.

\begin{figure}
	\includegraphics[width=0.5\textwidth]{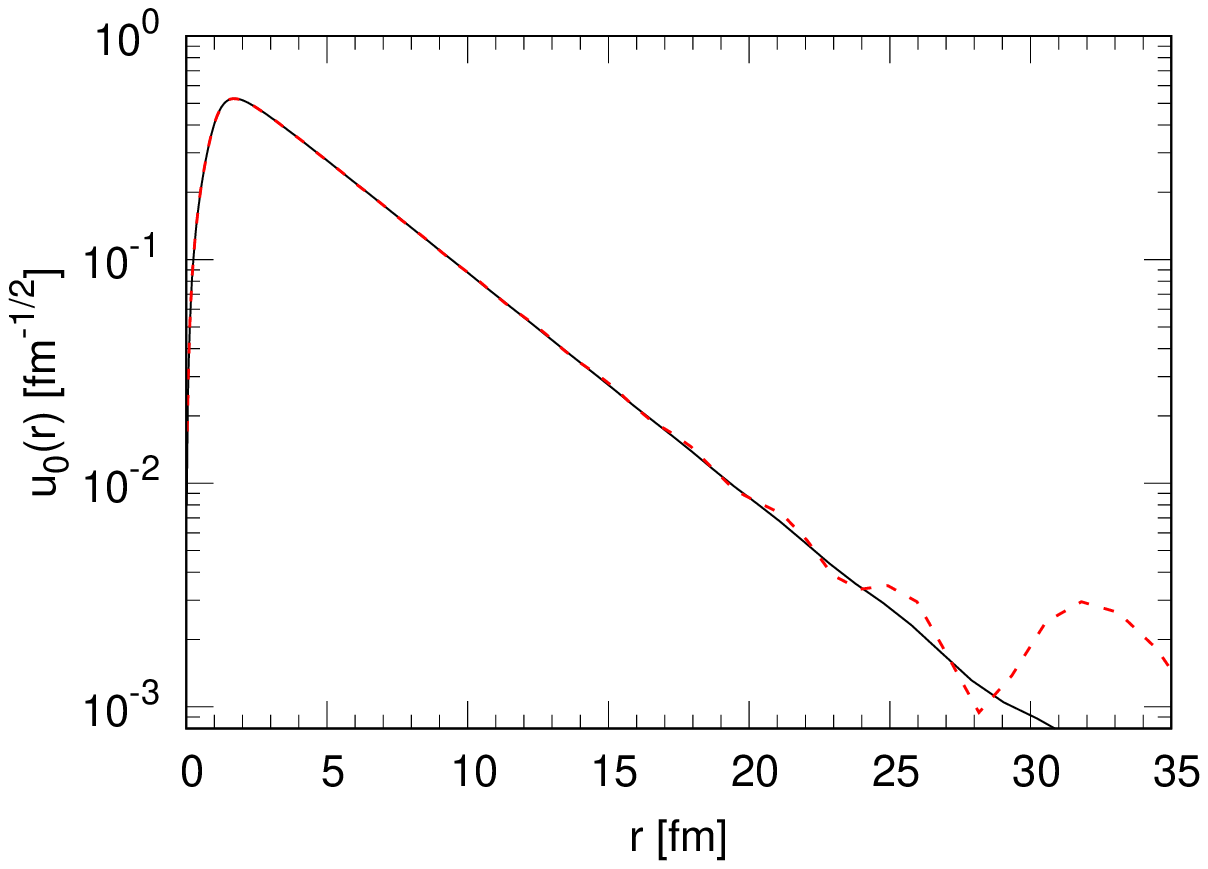}
	\includegraphics[width=0.5\textwidth]{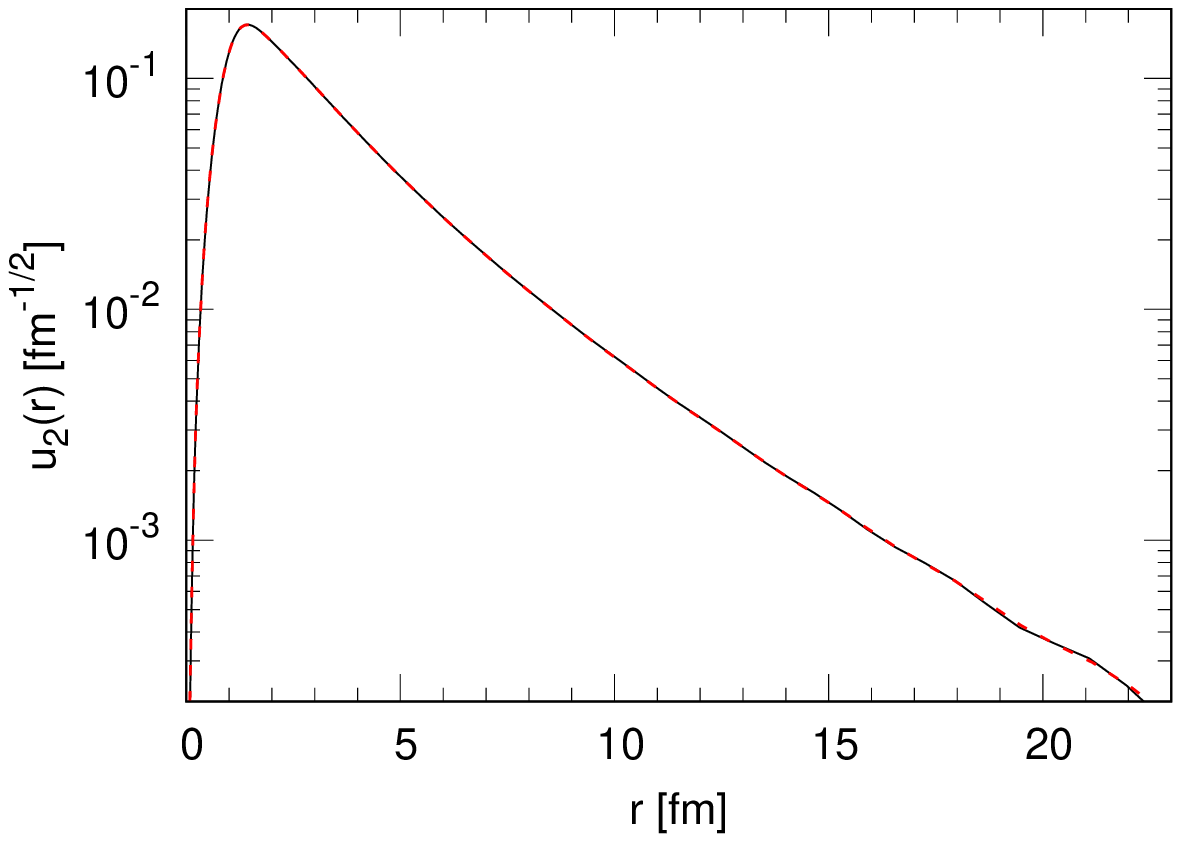}
    \includegraphics[width=0.5\textwidth]{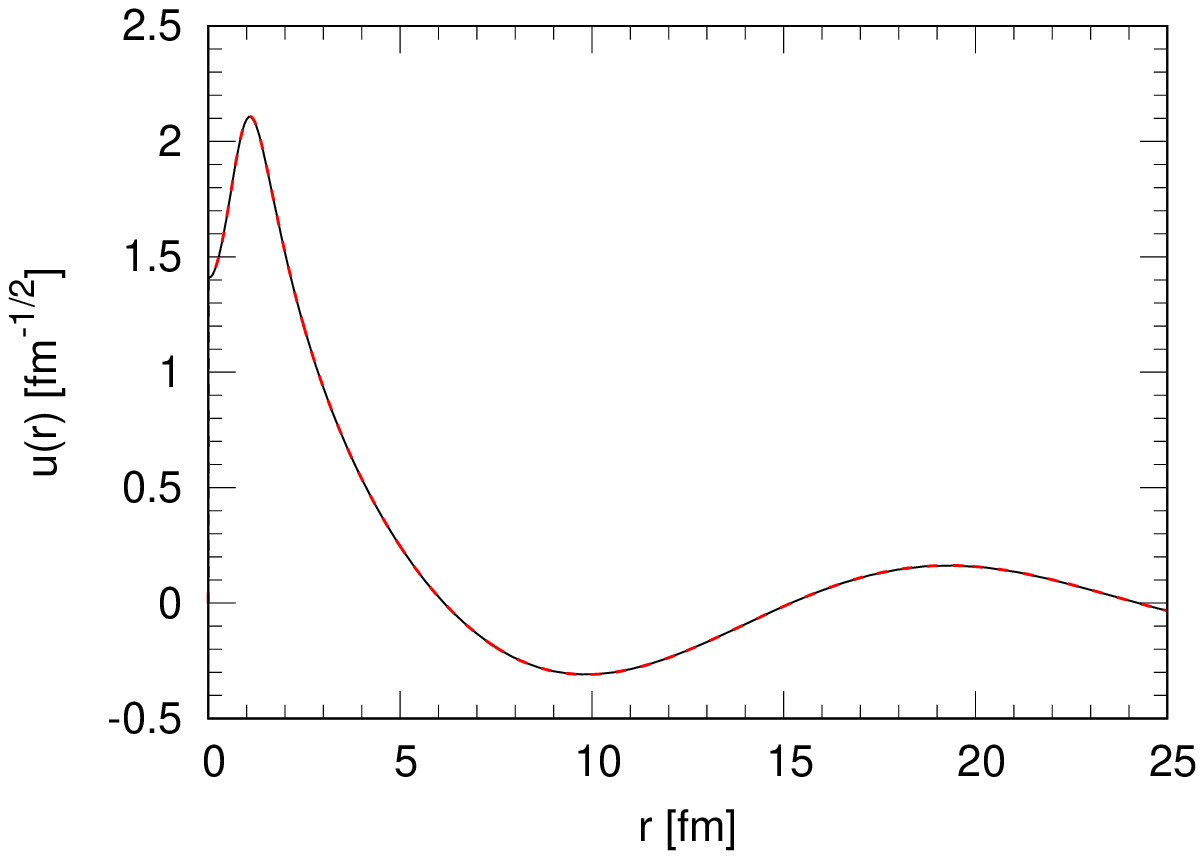}
        \caption{The deuteron
          $u_0(r)$ (left top panel) and $u_2(r)$ (right top panel)
          functions, and the $nn$ $^1S_0$ function (left bottom panel)
          at $E=5$ MeV are  
          calculated with the variational (dashed red line) and the
          Numerov (black line) methods. The NVIa potential is used.
          In order to appreciate the differences between
          the two methods, the function $u_0(r)$ and $u_2(r)$
          are shown in semilogarithmic scale.
          }
  \label{fig:A=2wfs}
\end{figure}

In order to compare the results obtained with the variational and the
Numerov methods, we report in Table~\ref{tab:be+ps} the deuteron binding
energies and the $nn$ phase shifts at the indicative
relative energy $E=5$ MeV
for the
four chiral potentials here under consideration. By inspection
of the table we can see an excellent agreement between the two
methods, with a difference well below 1 keV for the binding energies.
The phase shifts calculated with the two methods are as well in an
excellent numerical agreement.
Furthermore, we show in Fig.~\ref{fig:A=2wfs} the
deuteron and the $nn$ wave functions, still at $E=5$ MeV as an example,
for the NVIa potential. The results obtained with the other
chiral potentials present similar behaviour. By inspection of the
figure, we can see that the variational method fails to reproduce
the $u_0(r)$ function for $r>20$ fm. However, it should be noticed
that in this region, the function is almost two orders of magnitude
smaller than in the dominant range of $r\sim 0-5$ fm.
As we will see in the following section, we anticipate
already that these discrepancies
in the deuteron wave functions will have no impact on the
muon capture rate.
\begin{table}
  \begin{tabular}{c|cccc}
 Potential & $B_d$(Num.) & $B_d$(Var.) & $\delta_0$(Num.) & $\delta_0$(Var.) \\
\hline
 NVIa  & 2.22465 & 2.22464 & 57.714 & 57.714 \\
 NVIIa & 2.22442 & 2.22441 & 57.766 & 57.766 \\
 NVIb  & 2.22482 & 2.22486 & 57.815 & 57.812 \\
 NVIIb & 2.22418 & 2.22427 & 57.964 & 57.960 \\
  \end{tabular}
  \caption{Deuteron binding energies $B_d$, in MeV, and $nn$ $S$-wave
    phase shift $\delta_0$ at $E=5$ MeV, in deg, calculated with the Numerov (Num.)
    or the variational (Var.) methods using the four Norfolk
    chiral potentials NVIa, NVIIa, NVIb and NVIIb. Here we reports the results up to the digit from which
    the two methods start to differ. The experimental
  value for $B_d$ is $B_d^{exp}=2.2245$ MeV.}
  \label{tab:be+ps}
  \end{table}

\section{Results}
\label{sec:results}

We present in this section the results for the $\Gamma^D(^1S_0)$
muon capture rate, obtained using the Norfolk potentials and
consistent currents, as presented
in Sec.~\ref{subsec:pot-curr}. In particular, we will
use the four Norfolk potentials NVIa, NVIb, NVIIa, and NVIIb,
obtained varying the short- and long-range cutoffs (models a or b), and the
range of laboratory energies over which the fits have been carried out
(models I or II). For each model, the weak vector current and the axial
current and charge operators have been consistently constructed.
In particular, we will indicate with the label LO those results obtained
including only the LO contributions in the vector current
and axial current and charge operators, with NLO those ones obtained
including, in addition, the NLO contributions to the vector current and
axial charge operators. In fact, we remind that there are no NLO contributions
to the axial current. With the label N2LO
we will indicate those results obtained including the N2LO terms
of the vector and axial currents, but not the axial charge, since they vanish exactly. Finally, with N3LO we will
indicate the results obtained when N3LO terms in the vector and axial currents are retained. To be noticed that
this is the order at which new LECs appear. 
The contribution at N3LO for the axial charge are instead discarded for the reasons explained
in Sec.~\ref{subsec:pot-curr}.
 Finally,
we will use for the axial single-nucleon form factor the
dependence given in Eq.~(\ref{esp}) with $g_A=1.2723$ and
$r_A^2=0.46$ fm$^2$. However, in order to establish the uncertainty
arising from the rather poor knowledge of $r_A^2$ (see Ref.~\cite{Hill2018}
and the discussion in Sec.~\ref{sec:intro} and at the
end of Sec.~\ref{subsec:pot-curr}),
we will show also results obtained with
$r_A^2=0.30,0.46,0.62$ fm$^2$, so
that the 0.16 fm$^2$ uncertainty
on $r_A^2$~\cite{Hill2018} will be taken into account.

\begin{figure}
  \centering
	\includegraphics[width=0.6\textwidth]{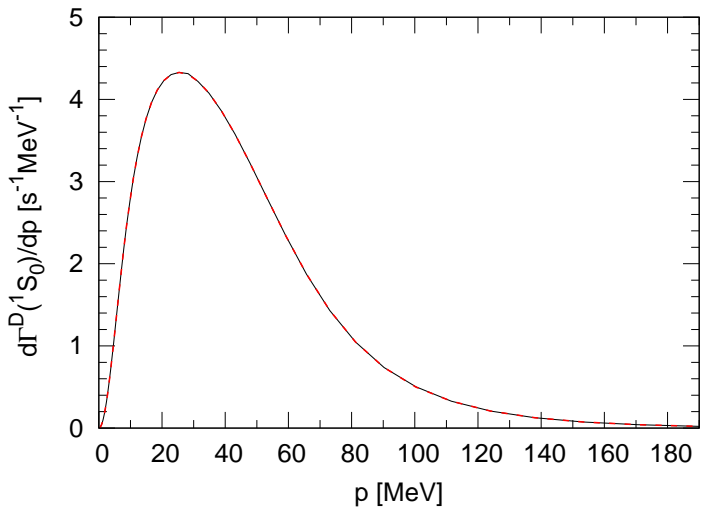}
  \caption{The differential doublet capture rate in the
    $^1S_0$ $nn$ channel, $d\Gamma^D(^1S_0)/dp$ in s$^{-1}$MeV$^{-1}$,
    as function of the $nn$ relative momentum $p$ in MeV,
    calculated using either the Numerov (black solid line)
    or the variational (red dashed line) methods in order
    to obtain the
    deuteron and the $nn$ scattering wave functions.
    The curves are exactly on the top of each other.
    The NVIa potential
    and consistent currents at N3LO are used. 
    The axial charge radius is taken to be $r_A^2=0.46$ fm$^2$.}
  \label{fig:nvia_var-num}
\end{figure}

Firstly, we begin by
proving that the uncertainty arising from the numerical method
adopted to study the deuteron and the $nn$ scattering states
is well below the 1\% level. In fact, in Table~\ref{tab:nvia_var-num}
we present the results obtained with the NVIa potential
and currents with up to N3LO contributions,
using either the variational
or the Numerov method to solve the two-body problem (see
Sec.~\ref{subsec:nuclwf}).
The function $d\Gamma^D(^1S_0)/dp$ (see Eq.~(\ref{eq:dgd})) calculated
with the same potential and currents 
is shown in Fig.~\ref{fig:nvia_var-num}. 
As it can be seen by inspection of the
figure and the table, the agreement between the results obtained
within the two methods is essentially perfect,
of the order of 0.01 s$^{-1}$ in $\Gamma^D(^1S_0)$, well below 
any other source of error ($\simeq0.005\%$).  Therefore,
from now on, we will present only results obtained using
the variational method, which is in fact numerically less involved
than the Numerov one.
\begin{table}
  \begin{tabular}{c|cc}
 $\chi$-order & Numerov & Variational \\
\hline
 LO   & 245.43 & 245.42 \\
 NLO  & 247.59 & 247.58 \\
 N2LO & 254.67 & 254.65 \\
 N3LO & 255.31 & 255.30 \\
  \end{tabular}
  \caption{
    The total doublet capture rate in the
    $^1S_0$ $nn$ channel, $\Gamma^D(^1S_0)$ in s$^{-1}$,
    calculated using either the Numerov or the variational methods
    to obtain the
    deuteron and the $nn$ scattering wave functions. Here we report the results up to the digit for which the two methods differ. The NVIa potential
    and consistent currents at the various chiral order are used,
    and the axial charge radius is taken to be $r_A^2=0.46$ fm$^2$.}
  \label{tab:nvia_var-num}
  \end{table}

We now present in Table~\ref{tab:fullresults} the results for
$\Gamma^D(^1S_0)$, obtained using 
all the four Norfolk potentials, NVIa, NVIb, NVIIa and NVIIb,
and consistent currents, from LO, up to N3LO. The axial
charge radius is fixed at $r_A^2=0.46$ fm$^2$.
By inspection of the table, we can provide our best estimate
for $\Gamma^D({^1S_0})$,
which we calculate simply as the average between the four
values at N3LO, $\Gamma^D(^1S_0)=$ 255.8 s$^{-1}$.
Furthermore, we would like to remark that the overall
model-dependence is quite small, the largest difference being
of the order of 1.1 s$^{-1}$ between the NVIa and NVIIb results, at
N3LO.
Going into more detail, (i) by comparing the NVIa (NVIIa) and
NVIb (NVIIb) results,
still at N3LO,
we can get a grasp on the cutoff dependence, which turns
out to be smaller than 1 s$^{-1}$ for both models I and II. (ii) By comparing
the NVIa (NVIb) and NVIIa (NVIIb) results, also in this
case at N3LO,
we can conclude that the dependence on the
$NN$ database used for the LECs fitting procedure in the potentials
is essentially of the same order. 
To remain conservative, we have decided to define the theoretical
uncertainty arising from model-dependence as the half range, i.e.\ 
\begin{equation}
  \Delta\Gamma^D(^1S_0)[{\rm mod-dep}]\equiv
  \frac{|\Gamma^D(^1S_0)_{\rm NVIIb}-\Gamma^D(^1S_0)_{\rm NVIa}|}{2}
  \label{eq:deltaG-mod} \ .
\end{equation}
From this we obtain $\Delta\Gamma^D(^1S_0)[{\rm mod-dep}]=0.6$ s$^{-1}$.

Still by inspection of Table~\ref{tab:fullresults}, we can conclude that the
chiral order convergence seems to be quite well under control for all the potential models. In fact,  in going from LO
to NLO, $\Gamma^D(^1S_0)$ has increased by 2.2 s$^{-1}$ for the a models, and 
2.5 s$^{-1}$ and 2.4 s$^{-1}$ for the models NVIb and NVIIb, respectively. This small change is due to the fact that the only correction appearing at NLO comes from the vector current.
Passing from NLO to N2LO  the muon capture rate increases of 7.1 s$^{-1}$ for the interactions NVIa and NVIIa, and 11.5 s$^{-1}$ and 11.3 s$^{-1}$
for the models NVIb and NVIIb, respectively. This can be understood considering that the
terms with the $\Delta$-isobar contributions appear at this order for the vector and axial current.
The convergence at N3LO shows instead a more involved behaviour:
for the models NVIa and NVIIa, $\Gamma^D(^1S_0)$ increase of  0.6 s$^{-1}$ and 0.9 s$^{-1}$ respectively while for the models NVIb and NVIIb the muon capture rate decreases of 3.5 s$^{-1}$ and 3.9 s$^{-1}$, respectively. Even if the results are in reasonable agreement with the expected chiral convergence behaviour (in particular for the models a), the chiral convergence of the current shows a significant dependence on the regularization, that we tracked back to the axial current corrections and in particular to the different value of the constant $c_D$ (see Section~\ref{subsec:pot-curr}). 
We find still remarkable that the results at N3LO obtained
with the various potentials, even if their
chiral convergence pattern are quite different, 
turn out to be within 1.1 s$^{-1}$.

The theoretical uncertainty
arising from the chiral order convergence of the nuclear weak transition operators can be studied using
the prescription of Ref.~\cite{Epelbaum2015}. Here we report the formula for the error at N2LO only. At this order, for each energy, we define the error for the differential capture rate (to symplify the notation from now on we use $\Gamma^D(p)=d\Gamma^D({}^1S_0)/dp$),  as
\begin{equation}
  \Delta\Gamma^D(p)\equiv \max \left\{
  Q^3|\Gamma^D_{\rm LO}(p)|,\,Q^2|\Gamma^D_{\rm NLO}(p)-\Gamma^D_{\rm LO}(p)|,\,
  Q|\Gamma^D_{\rm N2LO}(p)-\Gamma^D_{\rm NLO}(p)|\right\}
  \label{eq:deltagp} \ ,
\end{equation}
where we have assumed 
\begin{equation}
    Q=\frac{1}{\Lambda}\frac{p^8+m_\pi^8}{p^{7}+m_\pi^{7}}
\end{equation}
as in Ref.~\cite{Acharya2022} for the case of the $np\leftrightarrow d\gamma$ reaction. Here, $p$ is the relative momentum of the $nn$ system
and we assume a value of $\Lambda\simeq 550$ MeV, which is of the order of the 
cutoff of the adopted interactions. Analogous formula have been used to study the 
other orders (see Ref.~\cite{Epelbaum2015} for details).

\begin{figure}
  \centering
	\includegraphics[width=0.6\textwidth]{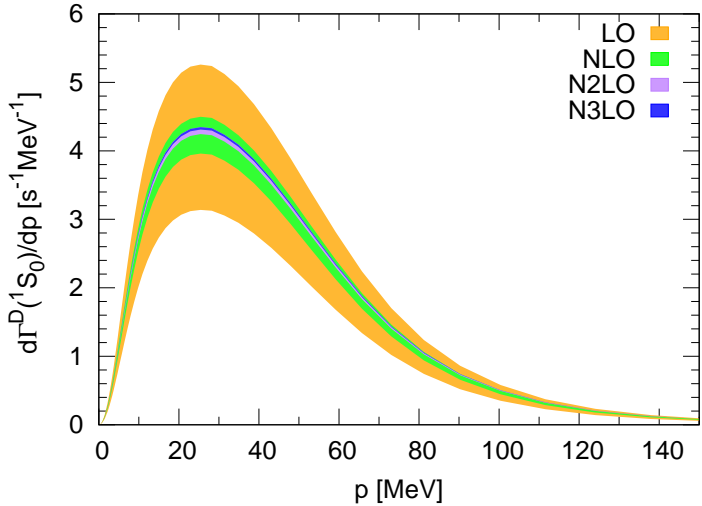}
  \caption{The differential doublet capture rate in the
    $^1S_0$ $nn$ channel, $d\Gamma^D(^1S_0)/dp$ in s$^{-1}$MeV$^{-1}$,
    as function of the $nn$ relative momentum $p$ in MeV,
    calculated order by order with the relative errors computed
    following the prescription of Ref.~\cite{Epelbaum2015}.
    The axial charge radius is taken to be $r_A^2=0.46$ fm$^2$.}
  \label{fig:nvia_curr-err}
\end{figure}

In Fig.~\ref{fig:nvia_curr-err} we show the error on $d\Gamma^D(^1S_0)/dp$ order by order in the expansion of the nuclear current up to N3LO for the NVIa interaction. From the figure it is evident the nice convergence of the chiral expansion.
The total error arising from the chiral truncation of the currents on $\Gamma^D(^1S_0)$ is then computed by integrating the error
of the differential capture rate over $p$, namely
\begin{equation}
  \Delta\Gamma^D({}^1S_0)[\text{curr-conv}]=\int_{0}^{p_{max}}\Delta\Gamma^D(p)dp\,.
\end{equation}
To be the most conservative possible we keep as error the largest obtained with the various interaction models. In the same spirit, we consider  the error computed at N2LO, since the calculation at N3LO does not contain all the contributions of the axial charge (see discussion Section~\ref{subsec:pot-curr}). Therefore we obtain $\Delta\Gamma^D({}^1S_0)[\text{curr-conv}]=4.4$
s$^{-1}$. In comparison the same calculation
at N3LO would give as error $\Delta\Gamma^D({}^1S_0)[\text{curr-conv}]|_{\rm N3LO}=1.4$ s$^{-1}$.
\begin{table}
  \begin{tabular}{c|cccc}
                 &\multicolumn{4}{c}{Potentials} \\
    $\chi$-order & NVIa & NVIb & NVIIa & NVIIb \\
\hline
 LO   & 245.4 & 245.1 & 245.7 & 246.6 \\
 NLO  & 247.6 & 247.6 & 247.9 & 249.0 \\
 N2LO & 254.7 & 259.1 & 255.0 & 260.3 \\
 N3LO & 255.3 & 255.6 & 255.9 & 256.4 \\
  \end{tabular}
  \caption{
    The total doublet capture rate in the
    $^1S_0$ $nn$ channel, $\Gamma^D(^1S_0)$ in s$^{-1}$,
    calculated using the four Norfolk potentials NVIa, NVIb,
    NVIIa, and NVIIb, and consistent currents, at the various chiral orders.
    The axial charge radius is taken to be $r_A^2=0.46$ fm$^2$,
    and the variational method is applied in order to
    calculate the deuteron and the $nn$ scattering wave functions.}
  \label{tab:fullresults}
  \end{table}

Finally, we present in Table~\ref{tab:res-ra} the results
obtained with all the interactions
and consistent currents up to N3LO 
for the three values of the axial charge radius, $r_A^2=0.30,0.46,0.62$ fm$^2$.
This will allow us to understand the importance of this
last source of theoretical uncertainty. The three values
have been chosen to span the range of values proposed in Ref.~\cite{Hill2018}.
Again, we define
the theoretical uncertainty $\Delta\Gamma^D(^1S_0)[r_A^2]$ 
arising from this last source
as the half range of the results, i.e.\
\begin{equation}
  \Delta\Gamma^D(^1S_0)[r_A^2]\equiv \max_{pot} \left\{
  \frac{|\Gamma^D(^1S_0)_{r_A^2=0.30}-\Gamma^D(^1S_0)_{r_A^2=0.62}|}{2}\right\}
  \label{eq:deltaG-ra} \ ,
\end{equation}
where with $\max_{pot}$ we indicate that we take the maximum value among the
different interactions considered.
By inspection of the table, we can conclude that $\Delta\Gamma^D(^1S_0)[r_A^2]=2.9$ s$^{-1}$,
which is found to be essentially model-independent.
\begin{table}
  \begin{tabular}{cccc}
     Pot. & $r_A^2=0.30$ & $r_A^2=0.46$ & $r_A^2=0.62$ \\
    \hline
    NVIa  & 258.2 & 255.3 & 252.4\\
    NVIb  & 258.5 & 255.6 & 252.8\\
    NVIIa & 258.7 & 255.9 & 253.0\\
    NVIIb & 259.3 & 256.4 & 253.6 \\
  \end{tabular}
  \caption{
    The total doublet capture rate in the
    $^1S_0$ $nn$ channel, $\Gamma^D(^1S_0)$ in s$^{-1}$,
    is calculated using all the different interactions
    and consistent currents up to N3LO, and three
    different values of $r_A^2$,
    $r_A^2=0.30,0.46,0.62$ fm$^2$.
    The variational method is applied in order to
    calculate the deuteron and the $nn$ scattering wave functions.}
  \label{tab:res-ra}
  \end{table}

In conclusion, our final result for $\Gamma^D(^1S_0)$ is
\begin{equation}
  \Gamma^D(^1S_0)=255.8(0.6)(4.4)(2.9)\,\, {\rm s}^{-1} \label{eq:final} \ ,
\end{equation}
where the three uncertainties arise from model-dependence, chiral
convergence and the experimental error in the axial charge radius $r_A$.
The overall systematic uncertainty becomes 5.0 s$^{-1}$,
when the various contributions are summed. The uncertainty on $r_A^2$ is instead a statistical uncertainty and therefore must be treated separately.
This result can be compared
with those of
Refs.~\cite{Marcucci2012chiral,Acharya2018}.
In Ref.~\cite{Marcucci2012chiral}, we found a value of 
$253.5(1.2)$ s$^{-1}$, the error taking care of the cutoff dependence
and the uncertainty in the
$d_R$ LEC fitting procedure. When only the cutoff
dependence is considered, it reduces to 0.2 s$^{-1}$, somewhat
smaller than the present 0.6 s$^{-1}$. The central values
that we have obtained and the one quoted in Ref.~\cite{Marcucci2012chiral},
even if the chiral potentials are very different, are instead in reasonable
agreement.
In Ref.~\cite{Acharya2018}, it was found
$\Gamma^D(^1S_0)=252.8(4.6)(3.9)\;\rm{s}^{-1}$, where the first
error is due to the truncation in the chiral expansion
and the second to the uncertainty in the nucleon axial radius 
$r_A$. These two errors should be compared with our $5.0$ s$^{-1}$ and
$2.9$ s$^{-1}$.
The agreement for the first error is very nice, while the
small difference in the second error is certainly due to the fact that
in Ref.~\cite{Acharya2018} a larger uncertainty for
$r_A^2$ was used ($0.22$ fm$^2$ vs.\ the present $0.16$ fm$^2$).
Also in this case, the agreement between the central values is 
good, even if the potential models adopted are very different.
This could suggest that the observable $\Gamma^D(^1S_0)$
is not sensitive to the nuclear
potential model, as long as this is able to
properly reproduce the deuteron and the $nn$ scattering
systems (as, in fact, any realistic modern potential
usually does).

\section{Conclusions and outlook}
\label{sec:concl}
We have investigated, for the first time with local nuclear potential models
derived in $\chi$EFT and consistent currents, the muon capture on deuteron,
in the $^1S_0$ initial $nn$ scattering state. The use of this framework
has allowed us to (i) provide a new estimate for the catpure rate
$\Gamma^D(^1S_0)$, which has turned out to be in agreement with the
results already present in the
literature and obtained still in
$\chi$EFT, but with different (non-local) potential
models~\cite{Marcucci2012chiral,Acharya2018};
(ii) accompany this estimate with a 
determination of the theoretical uncertainty, which arises
from model-dependence, chiral convergence, and the uncertainty
in the single-nucleon axial charge radius $r_A$.
We have also verified that the uncertainty
arising from the numerical technique adopted to solve
the two-body bound- and scattering-state problem
is completely negligible. This is in contrast with the conclusions
of Ref.~\cite{Acharya2017}, at least for the observable
$\Gamma^D(^1S_0)$. 

Our final result is $\Gamma^D(^1S_0)=255.8(0.6)(4.4)(2.9)$ s$^{-1}$, where
the three errors come from the three sources of uncertainty
just mentioned.
In order to provide an indicative value for the
overall uncertainty,
we propose to sum the systematic uncertainties arising from source (i)
and (ii), obtaining the value of 5.0 s$^{-1}$. Then, this error can be summed in quadrature
with the one of source (iii), 2.9 s$^{-1}$. Therefore,
we obtain $\Gamma^D(^1S_0)=255.8(5.8)$ s$^{-1}$. We remark again
that the value of 5.8 s$^{-1}$ for the overall uncertainty is only indicative, and the preferable procedure should be to treat the three
errors, 0.6 s$^{-1}$, 4.4 s$^{-1}$, and 2.9 s$^{-1}$, separately.

Given the success of this calculation in determining $\Gamma^D(^1S_0)$
and its uncertainty, with a procedure definitely less involved
than the one of Ref.~\cite{Acharya2018}, which still leads to similar
results, we plan to proceed applying this framework to the calculation
of $\Gamma^D$, retaining all the $nn$ partial waves up to $J=2$ and $L=3$.
These are known to provide contributions to $\Gamma^D$
up to 1 s$^{-1}$~\cite{Marcucci2011}.
In parallel, we plan to study the muon capture processes also on
$^3$He and $^6$Li, on the footsteps of Ref.~\cite{King2022a}. Here
the Norfolk potentials have been used in conjunction with the variational
and Green's
function Monte Carlo techniques to solve for the $A=3,6$ bound states, and the
final results have been found in some disagreement with the experimental
data. It will be interesting to verify these outcomes, using the
Hyperspherical Harmonics method
to solve for the $A=3,6$ nuclei~\cite{Marcucci2020,Gnech2020,Gnech2021}.
Last but not least, we plan to
apply this same framework to weak processes of interest for
Solar standard models and Solar neutrino fluxes, i.e.\ the
proton weak capture on proton (reaction~\ref{reaction2}), and
on $^3$He (the so called $hep$ reaction). In this second case,
it is remarkable that a consistent $\chi$EFT calculation is still missing (see Refs.~\cite{Marcucci2001,Park2003,Adelberger2011}).
For both reactions, we will
be able to provide a value for the astrophysical $S$-factor
at zero energy accompanied by an estimate of the
theoretical uncertainty.

\bibliographystyle{frontiersinHLTH&FPHY} 
\bibliography{mud-biblio}

\begin{thebibliography}{52}
\expandafter\ifx\csname natexlab\endcsname\relax\def\natexlab#1{#1}\fi
\expandafter\ifx\csname urlstyle\endcsname\relax
  \expandafter\ifx\csname doi\endcsname\relax
  \def\doi#1{doi:\discretionary{}{}{}#1}\fi \else
  \expandafter\ifx\csname doi\endcsname\relax
  \def\doi{doi:\discretionary{}{}{}\begingroup \urlstyle{rm}\Url}\fi \fi
\expandafter\ifx\csname selectlanguage\endcsname\relax
  \def\selectlanguage#1{}\fi

\bibitem[{Measday(2001)}]{Measday2001}
Measday DF.
\newblock The nuclear physics of muon capture.
\newblock {\em Phys. Rep.\/} {\bf 354} (2001) 243.

\bibitem[{Wang et~al.(1965)Wang, Anderson, Bleser, Lederman, Meyer, Rosen
  et~al.}]{Wang1965}
Wang IT, Anderson EW, Bleser EJ, Lederman LM, Meyer SL, Rosen JL, et~al.
\newblock Muon capture in (p $\mu$ d)+ molecules.
\newblock {\em Phys. Rev.\/} {\bf 139} (1965) B1528.

\bibitem[{Bertin et~al.(1973)Bertin, Vitale, Placci, and
  Zavattini}]{Bertin1973}
Bertin A, Vitale A, Placci A, Zavattini E.
\newblock Muon capture in gaseous deuterium.
\newblock {\em Phys. Rev. D\/} {\bf 8} (1973) 3774.

\bibitem[{Bardin et~al.(1986)Bardin, Duclos, Martino, Bertin, Capponi,
  Piccinini et~al.}]{Bardin1986}
Bardin G, Duclos J, Martino J, Bertin A, Capponi M, Piccinini M, et~al.
\newblock A measurement of the muon capture rate in liquid deuterium by the
  lifetime technique.
\newblock {\em Nucl. Phys. A\/} {\bf 453} (1986) 591.

\bibitem[{{Cargnelli M, et al.}(1989)}]{Cargnelli1989}
{Cargnelli M, et al}.
\newblock Workshop on fundamental $\mu$ physics, {L}os {A}lamos, 1986, {LA}
  10714{C}.
\newblock {\em Nuclear weak process and nuclear structure, Yamada Conference
  XXIII, edited by M. Morita, H. Ejiri, H. Ohtsubo, and T. Sato, World
  Scientific, Singapore\/}  (1989) 115.

\bibitem[{Kammel(2013)}]{Kammel2012}
Kammel P.
\newblock Precision muon capture at {PSI}.
\newblock {\em PoS CD12\/}  (2013) 016.

\bibitem[{Marcucci(2012)}]{Marcucci2012review}
Marcucci LE.
\newblock Muon capture on deuteron and $^3${He}: A personal review.
\newblock {\em Int. J. Mod. Phys. A\/} {\bf 27} (2012) 1230006.

\bibitem[{Adam et~al.(2012)Adam, Tater, Truhlik, Epelbaum, Machleidt, and
  Ricci}]{Adam2011}
Adam J Jr, Tater M, Truhlik E, Epelbaum E, Machleidt R, Ricci P.
\newblock {Calculation of Doublet Capture Rate for Muon Capture in Deuterium
  within Chiral Effective Field Theory}.
\newblock {\em Phys. Lett. B\/} {\bf 709} (2012) 93.

\bibitem[{Marcucci et~al.(2011)Marcucci, Piarulli, Viviani, Girlanda, Kievsky,
  Rosati et~al.}]{Marcucci2011}
Marcucci LE, Piarulli M, Viviani M, Girlanda L, Kievsky A, Rosati S, et~al.
\newblock {Muon capture on deuteron and $^3He$}.
\newblock {\em Phys. Rev. C\/} {\bf 83} (2011) 014002.

\bibitem[{Marcucci et~al.(2012)Marcucci, Kievsky, Rosati, Schiavilla, and
  Viviani}]{Marcucci2012chiral}
Marcucci LE, Kievsky A, Rosati S, Schiavilla R, Viviani M.
\newblock {Chiral effective field theory predictions for muon capture on
  deuteron and $^3$He}.
\newblock {\em Phys. Rev. Lett.\/} {\bf 108} (2012) 052502.
\newblock {[Erratum: {\it Phys. Rev. Lett.} {\bf 121}, (2018) 049901]}.

\bibitem[{Golak et~al.(2014)Golak, Skibi\'nski, Wita\l{}a, Topolnicki,
  Elmeshneb, Kamada et~al.}]{Golak2014}
Golak J, Skibi\'nski R, Wita\l{}a H, Topolnicki K, Elmeshneb AE, Kamada H,
  et~al.
\newblock {Break-up channels in muon capture on $^3$He}.
\newblock {\em Phys. Rev. C\/} {\bf 90} (2014) 024001.
\newblock [Addendum: Phys.Rev.C 90, 029904 (2014)].

\bibitem[{Acharya et~al.(2018)Acharya, Ekstr{\"o}m, and Platter}]{Acharya2018}
Acharya B, Ekstr{\"o}m A, Platter L.
\newblock Effective-field-theory predictions of the muon-deuteron capture rate.
\newblock {\em Phys. Rev. C\/} {\bf 98} (2018) 065506.

\bibitem[{Entem and Machleidt(2003)}]{Entem2003}
Entem DR, Machleidt R.
\newblock {Accurate charge dependent nucleon nucleon potential at fourth order
  of chiral perturbation theory}.
\newblock {\em Phys. Rev. C\/} {\bf 68} (2003) 041001.

\bibitem[{Machleidt and Entem(2011)}]{Machleidt2011}
Machleidt R, Entem DR.
\newblock {Chiral effective field theory and nuclear forces}.
\newblock {\em Phys. Rept.\/} {\bf 503} (2011) 1.

\bibitem[{Gazit et~al.(2009)Gazit, Quaglioni, and {Navr\'atil}}]{Gazit2009}
Gazit D, Quaglioni S, {Navr\'atil} P.
\newblock {Three-Nucleon Low-Energy Constants from the Consistency of
  Interactions and Currents in Chiral Effective Field Theory}.
\newblock {\em Phys. Rev. Lett.\/} {\bf 103} (2009) 102502.
\newblock {[Erratum: {\it Phys. Rev. Lett.} {\bf 122}, (2019) 029901]}.

\bibitem[{Carlsson et~al.(2016)Carlsson, Ekstr\"om, Forss\'en, Str\"omberg,
  Jansen, Lilja et~al.}]{Carlsson2016}
Carlsson BD, Ekstr\"om A, Forss\'en C, Str\"omberg DF, Jansen GR, Lilja O,
  et~al.
\newblock Uncertainty analysis and order-by-order optimization of chiral
  nuclear interactions.
\newblock {\em Phys. Rev. X\/} {\bf 6} (2016) 011019.

\bibitem[{Acharya et~al.(2017)Acharya, Ekstr\"om, Odell, Papenbrock, and
  Platter}]{Acharya2017}
Acharya B, Ekstr\"om A, Odell D, Papenbrock T, Platter L.
\newblock {Corrections to nucleon capture cross sections computed in truncated
  Hilbert spaces}.
\newblock {\em Phys. Rev. C\/} {\bf 95} (2017) 031301.

\bibitem[{Piarulli and Tews(2020)}]{Piarulli2020}
Piarulli M, Tews I.
\newblock Local nucleon-nucleon and three-nucleon interactions within chiral
  effective field theory.
\newblock {\em Front. Phys.\/} {\bf 7} (2020) 245.

\bibitem[{Piarulli et~al.(2016)Piarulli, Girlanda, Schiavilla, Kievsky, Lovato,
  Marcucci et~al.}]{Piarulli2016}
Piarulli M, Girlanda L, Schiavilla R, Kievsky A, Lovato A, Marcucci LE, et~al.
\newblock {Local chiral potentials with $\Delta$-intermediate states and the
  structure of light nuclei}.
\newblock {\em Phys. Rev. C\/} {\bf 94} (2016) 054007.

\bibitem[{Baroni et~al.(2018)Baroni, Schiavilla, Marcucci, Girlanda, Kievsky,
  Lovato et~al.}]{Baroni2018}
Baroni A, Schiavilla R, Marcucci LE, Girlanda L, Kievsky A, Lovato A, et~al.
\newblock Local chiral interactions, the tritium {G}amow-{T}eller matrix
  element, and the three-nucleon contact term.
\newblock {\em Phys. Rev. C\/} {\bf 98} (2018) 044003.

\bibitem[{Schiavilla et~al.(2019)Schiavilla, Baroni, Pastore, Piarulli,
  Girlanda, Kievsky et~al.}]{Schiavilla2019}
Schiavilla R, Baroni A, Pastore S, Piarulli M, Girlanda L, Kievsky A, et~al.
\newblock Local chiral interactions and magnetic structure of few-nucleon
  systems.
\newblock {\em Phys. Rev. C\/} {\bf 99} (2019) 034005.

\bibitem[{Gnech and Schiavilla(2022)}]{Gnech2022}
Gnech A, Schiavilla R.
\newblock {Magnetic structure of few-nucleon systems at high momentum transfers
  in a $\chi$EFT approach}  (2022).
\newblock {arXiv:2207.05528}.

\bibitem[{Piarulli et~al.(2018)}]{Piarulli2017}
Piarulli M, et~al.
\newblock {Light-nuclei spectra from chiral dynamics}.
\newblock {\em Phys. Rev. Lett.\/} {\bf 120} (2018) 052503.
\newblock \doi{10.1103/PhysRevLett.120.052503}.

\bibitem[{Gandolfi et~al.(2020)Gandolfi, Lonardoni, Lovato, and
  Piarulli}]{Gandolfi2020}
Gandolfi S, Lonardoni D, Lovato A, Piarulli M.
\newblock {Atomic nuclei from quantum Monte Carlo calculations with chiral EFT
  interactions}.
\newblock {\em Front. in Phys.\/} {\bf 8} (2020) 117.

\bibitem[{King et~al.(2020{\natexlab{a}})King, Andreoli, Pastore, Piarulli,
  Schiavilla, Wiringa et~al.}]{King2020a}
King GB, Andreoli L, Pastore S, Piarulli M, Schiavilla R, Wiringa RB, et~al.
\newblock {Chiral Effective Field Theory Calculations of Weak Transitions in
  Light Nuclei}.
\newblock {\em Phys. Rev. C\/} {\bf 102} (2020{\natexlab{a}}) 025501.

\bibitem[{King et~al.(2020{\natexlab{b}})King, Andreoli, Pastore, and
  Piarulli}]{King2020b}
King GB, Andreoli L, Pastore S, Piarulli M.
\newblock {Weak Transitions in Light Nuclei}.
\newblock {\em Front. in Phys.\/} {\bf 8} (2020{\natexlab{b}}) 363.

\bibitem[{King et~al.(2022{\natexlab{a}})King, Pastore, Piarulli, and
  Schiavilla}]{King2022a}
King GB, Pastore S, Piarulli M, Schiavilla R.
\newblock {Partial muon capture rates in A=3 and A=6 nuclei with chiral
  effective field theory}.
\newblock {\em Phys. Rev. C\/} {\bf 105} (2022{\natexlab{a}}) L042501.

\bibitem[{Cirigliano et~al.(2019)Cirigliano, Dekens, De~Vries, Graesser,
  Mereghetti, Pastore et~al.}]{Cirigliano2019}
Cirigliano V, Dekens W, De~Vries J, Graesser ML, Mereghetti E, Pastore S,
  et~al.
\newblock {Renormalized approach to neutrinoless double- $\beta$ decay}.
\newblock {\em Phys. Rev. C\/} {\bf 100} (2019) 055504.

\bibitem[{King et~al.(2022{\natexlab{b}})King, Baroni, Cirigliano, Gandolfi,
  Hayen, Mereghetti et~al.}]{King2022b}
King GB, Baroni A, Cirigliano V, Gandolfi S, Hayen L, Mereghetti E, et~al.
\newblock {Ab initio calculation of the $\beta$ decay spectrum of $^6$He}
  (2022{\natexlab{b}}).
\newblock ArXiv:2207.11179.

\bibitem[{Piarulli et~al.(2020)Piarulli, Bombaci, Logoteta, Lovato, and
  Wiringa}]{Piarulli2019}
Piarulli M, Bombaci I, Logoteta D, Lovato A, Wiringa RB.
\newblock {Benchmark calculations of pure neutron matter with realistic
  nucleon-nucleon interactions}.
\newblock {\em Phys. Rev. C\/} {\bf 101} (2020) 045801.

\bibitem[{Lovato et~al.(2022)Lovato, Bombaci, Logoteta, Piarulli, and
  Wiringa}]{Lovato2022}
Lovato A, Bombaci I, Logoteta D, Piarulli M, Wiringa RB.
\newblock {Benchmark calculations of infinite neutron matter with realistic
  two- and three-nucleon potentials}.
\newblock {\em Phys. Rev. C\/} {\bf 105} (2022).

\bibitem[{Acharya et~al.(2019)Acharya, Platter, and Rupak}]{Acharya2019}
Acharya B, Platter L, Rupak G.
\newblock {Universal behavior of $p$-wave proton-proton fusion near threshold}.
\newblock {\em Phys. Rev. C\/} {\bf 100} (2019) 021001.

\bibitem[{Marcucci et~al.(2013)Marcucci, Schiavilla, and
  Viviani}]{Marcucci2013}
Marcucci LE, Schiavilla R, Viviani M.
\newblock {Proton-Proton Weak Capture in Chiral Effective Field Theory}.
\newblock {\em Phys. Rev. Lett.\/} {\bf 110} (2013) 192503.
\newblock {[Erratum: {\it Phys. Rev. Lett.} {\bf 123}, (2019) 019901]}.

\bibitem[{Hill et~al.(2018)Hill, Kammel, Marciano, and Sirlin}]{Hill2018}
Hill RJ, Kammel P, Marciano WJ, Sirlin A.
\newblock Nucleon axial radius and muonic hydrogen—a new analysis and review.
\newblock {\em Rep. Prog. Phys.\/} {\bf 81} (2018) 096301.

\bibitem[{Walecka(1995)}]{Walecka1995}
Walecka J.
\newblock {\em Theorethical Nuclear and Subnuclear Physics\/} (London: Imperial
  College Press) (1995).

\bibitem[{Navarro~Pérez et~al.(2013)Navarro~Pérez, Amaro, and
  Ruiz~Arriola}]{Navarro2013}
Navarro~Pérez R, Amaro JE, Ruiz~Arriola E.
\newblock Coarse-grained potential analysis of neutron-proton and proton-proton
  scattering below the pion production threshold.
\newblock {\em Phys. Rev. C\/} {\bf 88} (2013) 064002.
\newblock {[Erratum: {\it Phys. Rev. C} {\bf 91}, (2015) 029901]}.

\bibitem[{Navarro~Pérez et~al.(2014{\natexlab{a}})Navarro~Pérez, E., and
  Ruiz~Arriola}]{Navarro2014a}
Navarro~Pérez R, E AJ, Ruiz~Arriola E.
\newblock Coarse grained $nn$ potential with chiral two-pion exchange.
\newblock {\em Phys. Rev. C\/} {\bf 89} (2014{\natexlab{a}}) 024004.

\bibitem[{Navarro~Pérez et~al.(2014{\natexlab{b}})Navarro~Pérez, E., and
  Ruiz~Arriola}]{Navarro2014b}
Navarro~Pérez R, E AJ, Ruiz~Arriola E.
\newblock Statistical error analysis for phenomenological nucleon-nucleon
  potentials.
\newblock {\em Phys. Rev. C\/} {\bf 89} (2014{\natexlab{b}}) 064006.

\bibitem[{Krebs et~al.(2007)Krebs, Epelbaum, and Meissner}]{Krebs2007}
Krebs H, Epelbaum E, Meissner U.
\newblock {Nuclear forces with $\Delta$ excitations up to
  next-to-next-to-leading order, part I: Peripheral nucleon-nucleon waves}.
\newblock {\em Eur. Phys. J. A\/} {\bf 32} (2007) 127.

\bibitem[{Baroni et~al.(2016)Baroni, Girlanda, Pastore, Schiavilla, and
  Viviani}]{Baroni2016}
Baroni A, Girlanda L, Pastore S, Schiavilla R, Viviani M.
\newblock Nuclear axial currents in chiral effective field theory.
\newblock {\em Phys. Rev. C\/} {\bf 93} (2016) 015501.

\bibitem[{Meyer et~al.(2016)Meyer, Betancourt, Gran, and Hill}]{Meyer2016}
Meyer AS, Betancourt M, Gran R, Hill RJ.
\newblock Deuterium target data for precision neutrino-nucleus cross sections.
\newblock {\em Phys. Rev. D\/} {\bf 93} (2016) 113015.

\bibitem[{Abramowitz and Stegun(1964)}]{abramowitz1964handbook}
Abramowitz M, Stegun IA.
\newblock {\em Handbook of mathematical functions with formulas, graphs, and
  mathematical tables\/}, vol.~55 (US Government printing office) (1964).

\bibitem[{Kohn(1948)}]{Kohn1948}
Kohn W.
\newblock Variational methods in nuclear collision problems.
\newblock {\em Phys. Rev.\/} {\bf 74} (1948) 1763.

\bibitem[{Johnson(1978)}]{Johnson1978}
Johnson BR.
\newblock The renormalized {N}umerov method applied to calculating bound states
  of the coupled-channel {S}chroedinger equation.
\newblock {\em J. Chem. Phys.\/} {\bf 69} (1978) 4678.

\bibitem[{Epelbaum et~al.(2015)Epelbaum, Krebs, and U.G.}]{Epelbaum2015}
Epelbaum E, Krebs H, UG M.
\newblock {Improved chiral nucleon-nucleon potential up to
  next-to-next-to-next-to-leading order}.
\newblock {\em Eur. Phys. J. A\/} {\bf 51} (2015) 53.

\bibitem[{Acharya and Bacca(2022)}]{Acharya2022}
Acharya B, Bacca S.
\newblock Gaussian process error modeling for chiral effective-field-theory
  calculations of $np\leftrightarrow d\gamma$ at low energies.
\newblock {\em Phys. Lett. B\/} {\bf 827} (2022) 137011.

\bibitem[{Marcucci et~al.(2020)Marcucci, Dohet-Eraly, Girlanda, Gnech, Kievsky,
  and Viviani}]{Marcucci2020}
Marcucci LE, Dohet-Eraly J, Girlanda L, Gnech A, Kievsky A, Viviani M.
\newblock {The Hyperspherical Harmonics method: a tool for testing and
  improving nuclear interaction models}.
\newblock {\em Front. in Phys.\/} {\bf 8} (2020) 69.

\bibitem[{Gnech et~al.(2020)Gnech, Viviani, and Marcucci}]{Gnech2020}
Gnech A, Viviani M, Marcucci LE.
\newblock {Calculation of the $^6$Li ground state within the hyperspherical
  harmonic basis}.
\newblock {\em Phys. Rev. C\/} {\bf 102} (2020) 014001.

\bibitem[{Gnech et~al.(2021)Gnech, Marcucci, Schiavilla, and
  Viviani}]{Gnech2021}
Gnech A, Marcucci LE, Schiavilla R, Viviani M.
\newblock {Comparative study of $^6$He~\ensuremath{\beta}-decay based on
  different similarity-renormalization-group evolved chiral interactions}.
\newblock {\em Phys. Rev. C\/} {\bf 104} (2021) 035501.

\bibitem[{Marcucci et~al.(2001)Marcucci, Schiavilla, Viviani, Kievsky, Rosati,
  and Beacom}]{Marcucci2001}
Marcucci LE, Schiavilla R, Viviani M, Kievsky A, Rosati S, Beacom JF.
\newblock {Weak proton capture on $^3$He}.
\newblock {\em Phys. Rev. C\/} {\bf 63} (2001) 015801.

\bibitem[{Park et~al.(2003)Park, Marcucci, Schiavilla, Viviani, Kievsky, Rosati
  et~al.}]{Park2003}
Park TS, Marcucci LE, Schiavilla R, Viviani M, Kievsky A, Rosati S, et~al.
\newblock {Parameter free effective field theory calculation for the solar
  proton fusion and hep processes}.
\newblock {\em Phys. Rev. C\/} {\bf 67} (2003) 055206.

\bibitem[{Adelberger et~al.(2011)}]{Adelberger2011}
Adelberger EG, et~al.
\newblock {Solar fusion cross sections II: the pp chain and CNO cycles}.
\newblock {\em Rev. Mod. Phys.\/} {\bf 83} (2011) 195.

\end{thebibliography}


\section*{Conflict of Interest Statement}
The authors declare that the research was conducted in the absence of any commercial or financial relationships that could be construed as a potential conflict of interest.

\section*{Author Contributions}

LC and LEM have shared the idea, the formula derivation and the computer code
implementation of the work presented here. LEM has also taken
the main responsibility
for the drafting of the manuscript. AG has contributed in reviewing the
codes and running them in order to obtain the final results presented here,
while
MP and MV have given valuable suggestions during the setting up of the
calculation. All the Authors have equally contributed in reviewing and
correcting the draft of the manuscript.

\section*{Funding}
The support by the U.S. Departmentof Energy, Office of Nuclear Science,
under ContractsNo. DE-AC05-06OR23177 is acknowledged by AG, while
the U.S. Department ofEnergy through the FRIB Theory Alliance Award No.
DE-SC0013617 is acknowledged by MP.

\section*{Acknowledgments}
The computational resources of the Istituto Nazionale di Fisica Nucleare
(INFN), Sezione di Pisa, are gratefully acknowledged. The final calculation was performed using resources of the National Energy Research Scientific Computing Center (NERSC), a U.S. Department of Energy Office of Science User Facility located at Lawrence Berkeley National Laboratory, operated under Contract No. DE-AC02-05CH11231.

\end{document}